\documentclass[a4paper,11pt]{article}
\pdfoutput=1
\usepackage{jcappub}
\usepackage{journal_abbrev}
\usepackage[svgnames]{xcolor}
\usepackage{xparse, xspace, verbatim, ulem, subcaption, float, comment, bm, bbold, xstring, placeins, graphicx, amsmath, amssymb, soul, hyperref, subcaption, adjustbox, siunitx, lmodern, enumitem, csquotes}
\xspaceaddexceptions{]\}}

\allowdisplaybreaks

\newcommand{\refeq}[1]{Eq.~(\ref{eq:#1})}          
\newcommand{\refeqs}[2]{Eqs.~(\ref{eq:#1})--(\ref{eq:#2})}          
          
\newcommand{\reffig}[1]{Fig.~\ref{fig:#1}}          
          
\newcommand{\refsec}[1]{Sec.~\ref{sec:#1}}          
\newcommand{\refapp}[1]{App.~\ref{app:#1}}
\newcommand{\reftab}[1]{Tab.~\ref{tab:#1}}

\def\be{\begin{equation}}
\def\ee{\end{equation}}
\def\bea{\begin{eqnarray}}
\def\eea{\end{eqnarray}}

\def\ba#1\ea{\begin{align}#1\end{align}}
\def\bab#1\eab{\begin{equation}\begin{aligned}[b]#1\end{aligned}\end{equation}}

\def\bg#1\eg{\begin{gather}#1\end{gather}}
\renewcommand{\>}{\right\rangle}

\renewcommand{\l}{\ell}

\newcommand{\CMB}{\mathrm{CMB}}
\newcommand{\ksz}{{\mathrm{kSZ}}}
\newcommand{\fgas}{f_{\mathrm{gas}}}

\newcommand{\obs}{{\mathrm{obs}}}
\newcommand{\instr}{\mathrm{instr}}
\newcommand{\beam}{\mathrm{beam}}
\newcommand{\eff}{\mathrm{eff}}
\newcommand{\Tcmb}{T_{\rm CMB}}
\newcommand{\kSZ}{\mathrm{kSZ}}
\newcommand{\LOS}{\mathrm{LOS}}
\newcommand{\Tobs}{T_{\mathrm{obs}}}

\newcommand{\Cl}[1]{C^{#1}_\l}

\newcommand{\zspec}{z_{\mathrm{spec}}}
\newcommand{\zphoto}{z_{\mathrm{photo}}}

\def\Msun{M_{\odot}}
\def\Msunh{h^{-1}M_{\odot}}

\newcommand{\s}{\sigma}

\renewcommand{\v}[1]{\bm{#1}}
\newcommand{\vg}[1]{\vec{#1}}
\newcommand{\vge}[1]{\overrightarrow{#1}}

\newcommand{\vs}{\v{s}}

\newcommand{\vx}{\v{x}}

\newcommand{\vv}{\v{v}}

\newcommand{\<}{\langle}
\renewcommand{\>}{\rangle}

\newcommand{\D}{\Delta}

\newcommand{\nhat}{\hat{n}}
\newcommand{\vnhat}{\v{\hat{n}}}

\newcommand{\eps}{\epsilon}

\newcommand{\vl}{\v{\l}}

\def\l{\ell}

\renewcommand{\L}{\Lambda}

\newcommand{\Om}{\Omega_\mathrm{m}}
\newcommand{\Ob}{\Omega_\mathrm{b}}
\newcommand{\Or}{\Omega_r}
\newcommand{\OK}{\Omega_K}
\newcommand{\OL}{\Omega_\Lambda}

\newcommand{\Mpch}{\,h^{-1}\;{\rm Mpc}}

\def\L{\Lambda}

\def\nhat{\hat{n}}
\def\vnhat{\hat{\v{n}}}

\def\N{\mathcal{N}}

\def\P{\mathcal{P}}

\def\C{\mathbf{C}}

\def\1{\mathbb{1}}

\def\emph#1{\textit{#1}}
\newcommand{\borg}{{\sc borg}}
\newcommand{\tcola}{{t\textsc{cola}}}
\newcommand{\smica}{{\texttt{SMICA}}}

\title{Taking measurements of the kinematic Sunyaev-Zel'dovich effect \emph{forward}: including uncertainties from velocity reconstruction with forward modeling}

\author[a]{Nhat-Minh Nguyen,}
\author[b]{Jens Jasche,}
\author[c]{Guilhem Lavaux, and}
\author[a]{Fabian~Schmidt}

\emailAdd{minh@mpa-garching.mpg.de}
\emailAdd{jens.jasche@fysik.su.se}
\emailAdd{guilhem.lavaux@iap.fr}
\emailAdd{fabians@mpa-garching.mpg.de}

\affiliation[a]{Max--Planck--Institut f\"ur Astrophysik,
  Karl--Schwarzschild--Stra\ss e 1, D--85748 Garching, Germany}
\affiliation[b]{The Oskar Klein Centre, Department of Physics, Stockholm University, Albanova University Center, SE 106 91 Stockholm, Sweden}
\affiliation[c]{Sorbonne Universit\'e, CNRS, UMR 7095, Institut d'Astrophysique de Paris, 98 bis bd Arago, 75014 Paris, France}


\abstract{
  We measure the kinematic Sunyaev-Zel'dovich (kSZ) effect, imprinted by maxBCG clusters, on the Planck SMICA map of the Cosmic Microwave Background (CMB). Our measurement, for the first time, directly accounts for uncertainties in the velocity reconstruction step through the process of Bayesian forward modeling. We show that this often neglected uncertainty budget typically increases the final uncertainty on the measured kSZ signal amplitude by $\simeq15\%$ at cluster scale. We observe evidence for the kSZ effect, at a significance of $\simeq2\sigma$. Our analysis, when applied to future higher-resolution CMB data, together with minor improvements in map-filtering and signal-modeling methods, should yield both significant and \emph{unbiased} measurements of the kSZ signal, which can then be used to probe and constrain baryonic content of galaxy clusters and galaxy groups.
  }

\begin{document}

\maketitle
\flushbottom


\section{Introduction}
\label{sec:intro}

Cosmic Microwave Background (CMB) photons, originated from the last-scattering surface, might encounter and interact with moving clouds of free electrons before reaching CMB telescopes or satellites.
This phenomenon leaves imprints of cosmic large-scale structures, specifically ionized electron gas inside clusters of galaxies, on the blackbody temperature anisotropies of the observed CMB, and is often referred to collectively as the Sunyaev-Zel'dovich (SZ) effect \cite{Sunyaev:1972, Ostriker:1986}.
At non-relativistic limit, the coherent part of the electron cloud's motion -- following its host cluster and the cosmic flow -- can be decoupled from the random thermal part. Since velocity of the former is typically small compared to the speed of light, in the electron gas rest frame $h\nu \ll m_ec^2$ and the photon-electron interaction can be well described by Thomson scattering. The Thomson scattering process essentially introduces a Doppler shift of CMB blackbody temperature in the comoving observer frame. This characteristic shift is specifically described as the \emph{kinematic} SZ effect.

Consider a single point source, located at position $\vx$ along direction $\v{\nhat}$ on the sky. Its kSZ signal is given by \cite{Phillips:1995, Birkinshaw:1999} -- who assumed the case of single, elastic scatterings and, as mentioned above, the regime of low-energy photons (Rayleigh-Jeans limit):
\be
\frac{\D T_\ksz (\v{\nhat})}{T_0} = -\s_T \int \mathrm{d}l\; \left(\frac{\v{v}_e (\vx)\cdot\v{\nhat}}{c}\right) n_e(\vx)
\label{eq:Tksz}
\ee
where $T_0=2.725\times10^{6}\,\mu K$ is the CMB blackbody temperature, $n_e(\vx)$ is the free electron number density at position $\vx$, and $\v{v}_e(\vx)$ is the peculiar velocity of free electrons, while $\s_T$ and $c$ denote the Thomson scattering cross-section and the speed of light in vacuum, respectively. The integral $\int dl$ is performed along the line-of-sight (LOS) $\v{\nhat}$, from the detector to the last-scattering surface.
It is generally assumed that the bulk motion of galaxy clusters follows the large-scale motion of dark matter (DM) \cite{Desjacques:2018}, i.e. $\vv_{e}=\vv_{DM}=\vv$, and since the correlation length of the latter is much larger than the physical size of a typical galaxy cluster \cite{Planck:2016a}, \refeq{Tksz} could be further simplified to
\be
\frac{\D T_\ksz (\v{\nhat})}{T_0} = - \tau(\vx, \v{\nhat}) \left(v^{\mathrm{LOS}}(\vx, \v{\nhat})/c\right),
\label{eq:Tksztau}
\ee
where $v^{\LOS}(\vx, \vnhat)\equiv \vv(\vx) \cdot \vnhat$ denotes the velocity along the LOS $\vnhat$, and we have defined
\be
\tau(\vx, \vnhat) = \s_T\int \mathrm{d}l\; n_e(\vx)
\label{eq:taudef}
\ee
to be the LOS projected optical depth.

The right-hand side (r.h.s) of \refeq{Tksztau} indicates that measurements of the kSZ signal constrain the product $(\v{v}(\vx)\cdot\v{\nhat})\tau$.
Assuming an external constraint on $\tau$ (see, for example, \cite{Madhavacheril:2019}), the kSZ signal
then directly measures the peculiar velocity field $\v{v}(\vx)$ and hence allows for constraints not only on modified gravity and Dark Energy models \cite{Keisler:2013, Ma:2014, Mueller:2015a} but also the sum of neutrino masses \cite{Mueller:2015b}.
Turning \refeq{Tksztau} the other way around, given $\v{v}(\vx)$, say, reconstructed from galaxy survey data, the kSZ signal directly probes the optical depth $\tau$.
The kSZ signal, as can be seen from \refeq{Tksztau}, does not depend on the gas temperature and scales \emph{linearly} with the gas density, similar to the scaling with the gas velocity.
Because of that, late-time contribution to the kSZ effect from collapsed, virialized objects appear to be the perfect candidate for probing the otherwise elusive \emph{Warm-Hot Intergalactic Medium} (WHIM) \cite{Hernandez:2015, Lim:2017}. This diffuse form of free baryonic gas -- whose typical temperature is of order $10^{5}-10^{7}K$ -- is too cold to show up in X-ray and thermal SZ measurements. There are mounting evidences suggesting that WHIM might host a large fraction of baryons in our Universe that are still missing compared to the number predicted by our standard cosmological model \cite{Fukugita:1998, Shull:2012, Lim:2017, Nicastro:2018}.
Further, early contributions from the epoch of reionization \cite{Hu:2000,Dvorkin:2009} can be used to add an additional constraint on the optical depth at re-ionization $\tau_{\mathrm{rei}}$, which would then break the degeneracy between it and the amplitude of the primordial power spectrum $A_s$ in CMB measurements \cite{Alvarez:2020}.
Both directions will undoubtedly benefit from upcoming high-resolution CMB experiments and large-volume galaxy redshift surveys. Indeed, the advent of CMB-S4 \cite{CMB-S4:2016} should allow for the novel application of kSZ tomography, i.e. measurements of kSZ signal at different redshifts. The result of which could then be cross-correlated with datasets from DESI \cite{DESI:2013} or LSST \cite{LSST:2019} to constrain either primordial, local non-Gaussianity \cite{Munchmeyer:2019} or the evolution of ionized gas \cite{Ma:2017}.

This wealth of scientific return from kSZ measurements has motivated several attempts to detect this effect using various datasets and estimators, despite the fact that the kSZ signal is deeply buried beneath the primary CMB anisotropies.
These efforts have, against the odds, resulted in $\simeq 2-4\sigma$ evidence of the kSZ effect using the kSZ \emph{pairwise} estimator \cite{Hand:2012, Planck:2016a, Soergel:2016, Bernardis:2017, Sugiyama:2018} and the cross-correlation between CMB maps and a reconstructed velocity field \cite{Planck:2016a, Hernandez:2015, Schaan:2016, Lim:2017, Tanimura:2020}. In this paper, we follow the second approach.

Previously published implementations of this approach \citep[e.g.][]{Planck:2016a, Hernandez:2015, Schaan:2016, Lim:2017, Tanimura:2020} relied on a reconstruction of the peculiar velocity field around clusters, where $\v{v}(\vx)$ is derived -- assuming a certain cosmology with Hubble parameter $H$ and cosmic linear growth rate $f=d\ln\delta/d\ln a$ -- by solving the inversion of the \emph{linearized} continuity equation in either real-space $\vx$ \cite{Planck:2016a, Hernandez:2015, Lim:2017, Tanimura:2020}
\be
\v{\nabla} \cdot \v{v}(\vx) = -aHf\delta(\vx),
\label{eq:inverted_continuity}
\ee
or redshift-space $\vs$ \cite{Schaan:2016},
\be
\v{\nabla} \cdot \v{v}(\vs) + f \v{\nabla}\cdot \left[\left(\v{v}(\vs)\cdot\v{\nhat}\right)\v{\nhat}\right] = -aHf\delta(\vs),
\label{eq:inverted_continuity_redshift_space}
\ee
where the DM density field $\delta(\vx)$ is simply obtained from a smoothed galaxy density field $\delta_g(\vx)$ by assuming a \emph{local}, \emph{linear} bias relation of the form $\delta_g(\vx)=b_1\delta(\vx)$.
Specifically, Ref.~\cite{Planck:2016a, Hernandez:2015} and Ref.~\cite{Lim:2017} used galaxy and galaxy group catalogs extracted from SDSS-DR7 \cite{Abazajian:2008}, while Ref.~\cite{Schaan:2016} and Ref.~\cite{Tanimura:2020} used CMASS and both CMASS, LOWZ galaxy catalogs obtained from SDSS-DR11 \cite{Anderson:2013} and SDSS-DR12 \cite{Reid:2015}.
This simple method, however, yields only one single realization of the velocity field, regardless of uncertainties in the observed galaxy field. The resulting velocity field thus includes systematics that can potentially bias the kSZ detection and measurement, and the quoted uncertainty on the kSZ signal does not contain the propagated error from the uncertainty in the reconstructed velocity.

In this paper, we derive a posterior for the kSZ signal -- as filtered from a temperature anisotropy map at locations of massive clusters -- accounting for uncertainties in the velocity reconstruction process by \emph{marginalizing over an ensemble} of realizations of this field, all of which are compatible with the observed distribution of galaxies \citep[see, e.g.][]{Jasche:2013,Jasche:2019}. We apply our method on the Planck \smica2018 CMB map \cite{Planck:2018IV} and the maxBCG cluster catalog \cite{Koester:2007}. Our ensemble of velocity reconstruction is obtained from the Bayesian forward modeling of galaxy clustering using LOWZ and CMASS galaxy samples presented in \cite{Lavaux:2019}.

For consistency, except for the estimation of the CMB contribution to the covariance matrix in \refsec{kSZ:multi_scale_model}, in this paper we assume the same flat $\Lambda$CDM cosmology assumed by the reconstruction in \cite{Lavaux:2019}, with $\Or=0$, $\OK=0$, $\Om=0.2889$, $\Ob=0.048597$, $\OL=0.7111$, $w=-1$, $n_s=0.9667$, $\sigma_8=0.8159$, $H_0=67.74\,\mathrm{km} \mathrm{s}^{-1} \mathrm{Mpc}^{-1}$.
The paper is structured as follows. In section \refsec{kSZ:dataset}, we describe the datasets used in this work. After formulating the physical models and statistical methods to be applied on the data in section \refsec{kSZ:datamodel}, we report our measurements of the kSZ effect from maxBCG clusters and their associated uncertainties in section \refsec{kSZ:result}. We then assert the robustness and significance of our measurement by means of different null tests in \refsec{kSZ:test}. Finally, we summarize our results and discuss relevant systematics as well as future improvements of kSZ measurement in section \refsec{kSZ:conclusion}.


\section{Data}
\label{sec:kSZ:dataset}

\subsection{CMB data}
\label{sec:kSZ:cmb_data}

We recap here the main features of our CMB data, the Planck \smica{} temperature (intensity) map, and subsequently, describe our method of extracting a noisy estimate of the kSZ signal induced by a given galaxy cluster from this map.

\subsubsection{Planck \smica{} CMB map}
\label{sec:kSZ:smica_map}
Our choice of CMB data is the \smica{} temperature map from the Planck 2018 release\footnote{\url{https://pla.esac.esa.int/}} \cite{Planck:2018IV} (\smica2018 hereafter).
\smica{} (Spectral Matching Independent Component Analysis) \cite{Cardoso:2008} linearly combines Planck frequency channels with multipole-dependent weights, including multipoles up to $\l=4000$ \cite{Planck:2018IV}, into a single CMB intensity map.

Due to the finite resolution and detector noise associated with any CMB instrument, the observed temperature anisotropy $\D \Tobs$ is a convolution of the true anisotropy $\D T$ -- including both primary, i.e. primordial CMB, and secondary anisotropies, e.g. kSZ, thermal SZ (tSZ), integrated Sachs-Wolfe, etc. -- with the instrumental beam function $B$\footnote{Strictly speaking, there is also a convolution with the pixel window function of the HEALPix map, which we include in our analysis but omit in the equations here and below, for readability.}, plus the instrumental noise $\D T_{\instr}$, i.e.
\be
\D \Tobs (\v{\theta_i},\v{\theta}) = \int d^2\theta' \, \D T (\v{\theta_i},\v{\theta'})\, B(\v{\theta}_i,\v{\theta'})
+ \D T_{\instr}(\v{\theta_i},\v{\theta}),
\label{eq:kszobs}
\ee
where we have assumed the \emph{flat-sky} approximation and replaced the three-dimensional vectors $\vx$ and $\vnhat$ by the two-dimensional vector $\v{\theta_i}$ for a specific cluster $i$.
The effective beam function of the \smica2018 intensity map can be well approximated by a spherically symmetric Gaussian function
\be
B(\v{\theta}_i,\v{\theta}') = \frac{1}{\sqrt{2\pi\theta^2_{\beam}}} \exp{\left[-\frac{|\v{\theta}'-\v{\theta}_i|^2}{2\theta^2_{\beam}}\right]}
\ee
where the 5-arcmin full-width-half-maximum (FWHM) resolution translates into $\theta_{\beam}\approx5.0\,\mathrm{arcmin}/\sqrt{8\ln(2)}\approx2.1\,\mathrm{arcmin}$.

\subsubsection{Signal extraction}
\label{sec:kSZ:signal_extraction}
To extract the kSZ signal from a CMB map, an aperture photometry (AP) filter of radius $\theta_f$ is applied at the location of all clusters.
The extracted flux $\D T^{\theta_f}$ can then be expressed as a convolution of the observed flux  $\D \Tobs$ with a radial weight function $W^{\theta_f}$ associated with that AP filter, i.e.
\be
\D T^{\theta_f}(\v{\theta_i}) = \int d^2\theta\,W^{\theta_f}(\v{\theta}-\v{\theta_i})\, \D \Tobs(\v{\theta_i},\v{\theta})
\label{eq:kszap}
\ee
Specifically, the spherically symmetric weight function $W^{\theta_f}$ is given by
\be
W^{\theta_f}(\v{\theta}-\v{\theta_i})= W^{\theta_f}\left(|\v{\theta}-\v{\theta_i}|\right) = \begin{cases}
   1  & 0\leq|\v{\theta}-\v{\theta_i}|<\theta_f                \\
   -1 & \theta_f\leq|\v{\theta}-\v{\theta_i}|<\sqrt{2}\theta_f \\
   0  & \rm otherwise.
\end{cases}
\label{eq:apfilter}
\ee
This \emph{compensated} filter is designed to reduce contributions from primary CMB anisotropies, as well as other low-redshift sources of contamination\footnote{Contributions from structures below the redshift range of maxBCG clusters and the \borg-SDSS3 volume, as well as CMB foregrounds, generally manifest themselves as large-scale anisotropies in the observed CMB.}, which vary on scales larger than $\theta_f$ in the extracted flux. Thus, as $\theta_f$ increases, so does contamination from these sources in the filtered flux.
Consequently, the signal would be underestimated at small filter sizes, where parts of the signal fall outside the inner disks and are thus subtracted out. Once the whole cluster is encompassed by the filter, the signal should asymptote to a limiting value\footnote{The exact value of this asymptotic limit depends on various factors, it however should be proportional to the free baryonic fraction within the clusters \cite{Schaan:2016}.} while the uncertainty should increase due to increasing contamination.

 In our analysis, we measure the kSZ signal while varying the filter size $\theta_f$.
Specifically for the individual-scale measurements of the kSZ signal in \refsec{kSZ:individual_scale_model}, \refsec{kSZ:individual_scale_AAP_result}, and \refsec{kSZ:individual_scale_AAP_test}, we adopt an \emph{adaptive} aperture photometry (AAP) filter whose radius $\theta_{f,i}$ scales with the effective apparent size of cluster $i$.
This ensures that the filter always probes the same fraction of baryonic gas for each cluster -- assuming a universal gas profile.
In practice, the application of an AP or AAP filter amounts to taking the difference between the pixel-averaged temperature anisotropies within the inner disk and that within the outer ring.
For this estimate of the kSZ signal, the primary noise source for large filter sizes is still the primary CMB, while for small filter sizes -- where only a very limited number of pixels are encompassed by the inner disk or outer ring -- the instrumental noise dominates.

Our method of extracting the kSZ flux in this paper is similar to that in \cite{Hernandez:2015, Planck:2016a, Schaan:2016, Sugiyama:2018, Soergel:2018}. It is worth mentioning here that the typical apparent size of maxBCG clusters selected for our analysis is very close to the Planck beam (see \refsec{kSZ:cmb_data} and \refsec{kSZ:maxbcg_catalog}). We defer a more optimal filtering method (which would require more specific assumptions on the form of the gas profile), such as the matched filter \citep[see, e.g.][]{Li:2014, Soergel:2016, Lim:2017}, to applications on CMB data with higher resolutions.


\subsection{Galaxy cluster data}
\label{sec:kSZ:cluster_data}

In this section, we first review relevant properties of our galaxy cluster data, the maxBCG cluster catalog. We then detail how we model the cluster optical depth.

\subsubsection{MaxBCG cluster catalog}
\label{sec:kSZ:maxbcg_catalog}
The public version of maxBCG catalog, a \emph{volume-limited}, red-sequence galaxy cluster sample, includes clusters identified from the SDSS data. These clusters span a scaled richness $N_{200}=10-188$ and a redshift range of $z=0.1-0.3$ \cite{Koester:2007}.
We use the mean richness-mass relation given by Eq. (A15) in \cite{Rozo:2009} to convert the scaled richness $N_{200}$ of maxBCG clusters into $M_{200}$, the cluster total mass within the $R_{200}$ radius, accounting for the difference in mass definitions and cosmology \citep[see][appendix and references therein for details]{Johnston:2007}.
If we assume that the projected gas distribution in cluster $i$, with physical size $R_{200,i}$ and angular diameter distance $D_{A,i}$, can be approximated by a Gaussian profile (cf. \refeq{integrated_tau}), then the width of its profile -- in the flat-sky approximation -- is given by
\be
\theta_{\eff, i} = \sqrt{\theta^2_{200, i} + \theta^2_{\beam}}
\ee
where $\theta_{200,i}=R_{200,i}/D_{A,i}$.
As mentioned earlier in \refsec{kSZ:signal_extraction}, for the case of AAP filter, the radius of each filter is adapted to each specific cluster as $\theta_f=\varphi_f\,\theta_{200, i}$.
It is worth pointing out that an overall shift in the $N_{200}-M_{200}$ relation would affect only the signal amplitude but neither the significance nor the signal-to-noise (S/N) of the kSZ detection.
The catalog mean cluster mass and redshift are $M_{200}=1.288\times10^{14}\Msun$ and $z=0.23$ respectively. Below, we describe various selection cuts that we apply on the original maxBCG catalog, and the resulting sub-sample of maxBCG clusters used in our analysis.

Firstly, to avoid any possible tSZ contamination, we exclude clusters whose $M_{200}>0.85\times10^{14}\Msun$ from our analysis (see \refapp{kSZ:masscut}).
Next, we select only clusters within regions where the \borg-SDSS3 reconstruction are well-constrained by data, i.e. sky regions where LOWZ and CMASS galaxies are actually observed.
In addition, we remove clusters outside of the Planck 2018 common confidence mask recommended for temperature analysis \cite{Planck:2018IV}.

\begin{figure}
    \centerline{\resizebox{0.7\textwidth}{!}{\includegraphics*{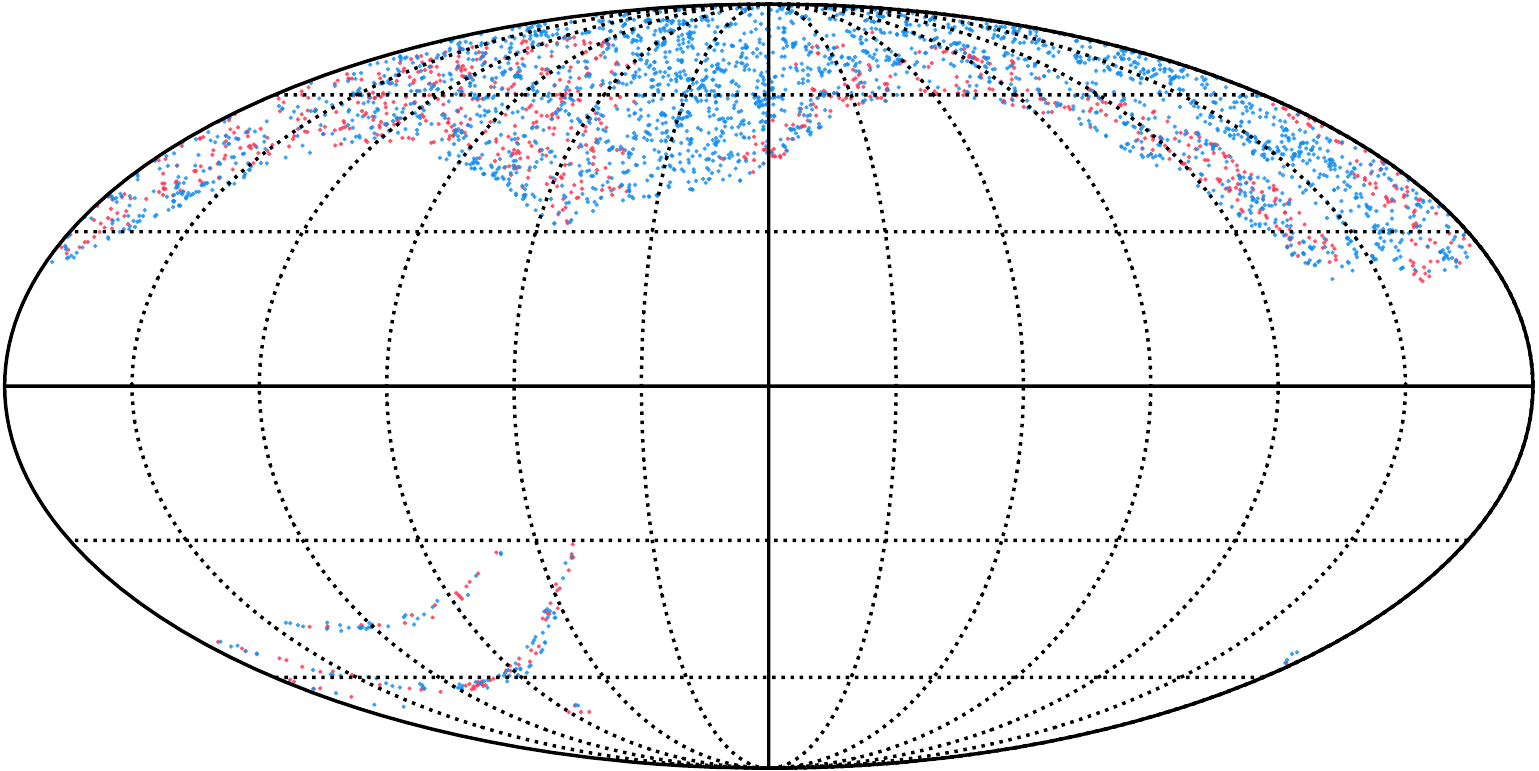}
    }}
    \caption{HEALPix projected map of 3512 clusters selected for our analysis, with 908 clusters with BCG spectroscopic redshift in red circles and 2604 clusters with photometric redshift in blue circles. The size of the circles are scaled with $\log_{10}(M_{200})$}\label{fig:selected_maxBCG_clusters_HEALPix_map}
  \end{figure}

This leaves us with a final sub-sample consisting of 3512 clusters from the original maxBCG catalog. We show in \reffig{selected_maxBCG_clusters_HEALPix_map} a HEALPix map\footnote{\url{https://healpix.sourceforge.io}} \cite{Gorski:2005,Zonca:2019} of these clusters in Galactic coordinates. We further divide our cluster sample into two datasets:
\begin{enumerate}[leftmargin=1.5cm]
   \item[spec-z:] this set includes 908 clusters whose brightest member galaxies (BCG) have spectroscopically determined redshifts, denoted by $\zspec$;
   \item[photo-z:] this set includes 2604 clusters for which only photometric redshifts as inferred by the maxBCG algorithm, denoted by $\zphoto$, are available.
\end{enumerate}

\begin{figure*}
    \centerline{\resizebox{\hsize}{!}{\includegraphics*{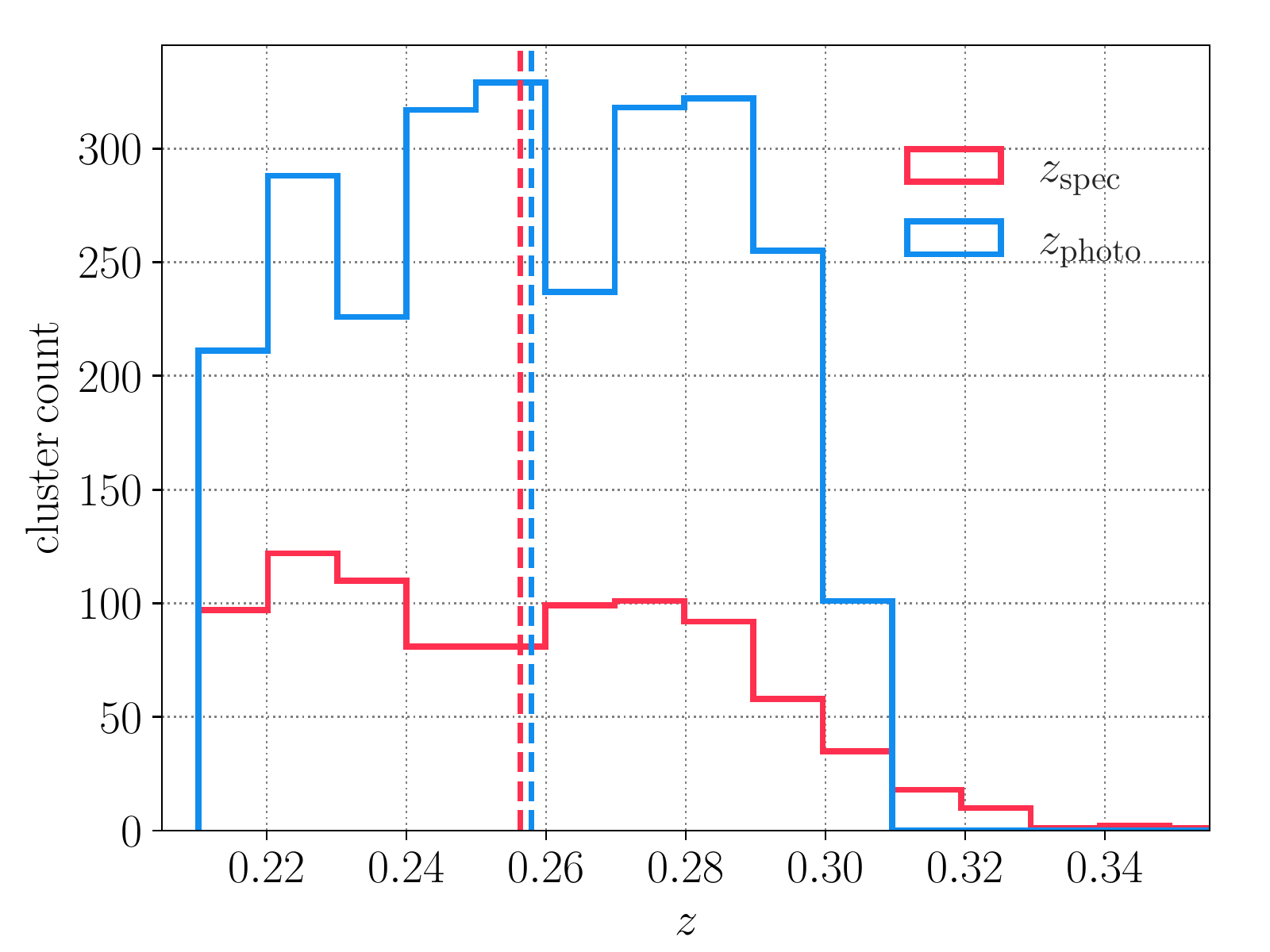} \,
        \includegraphics*{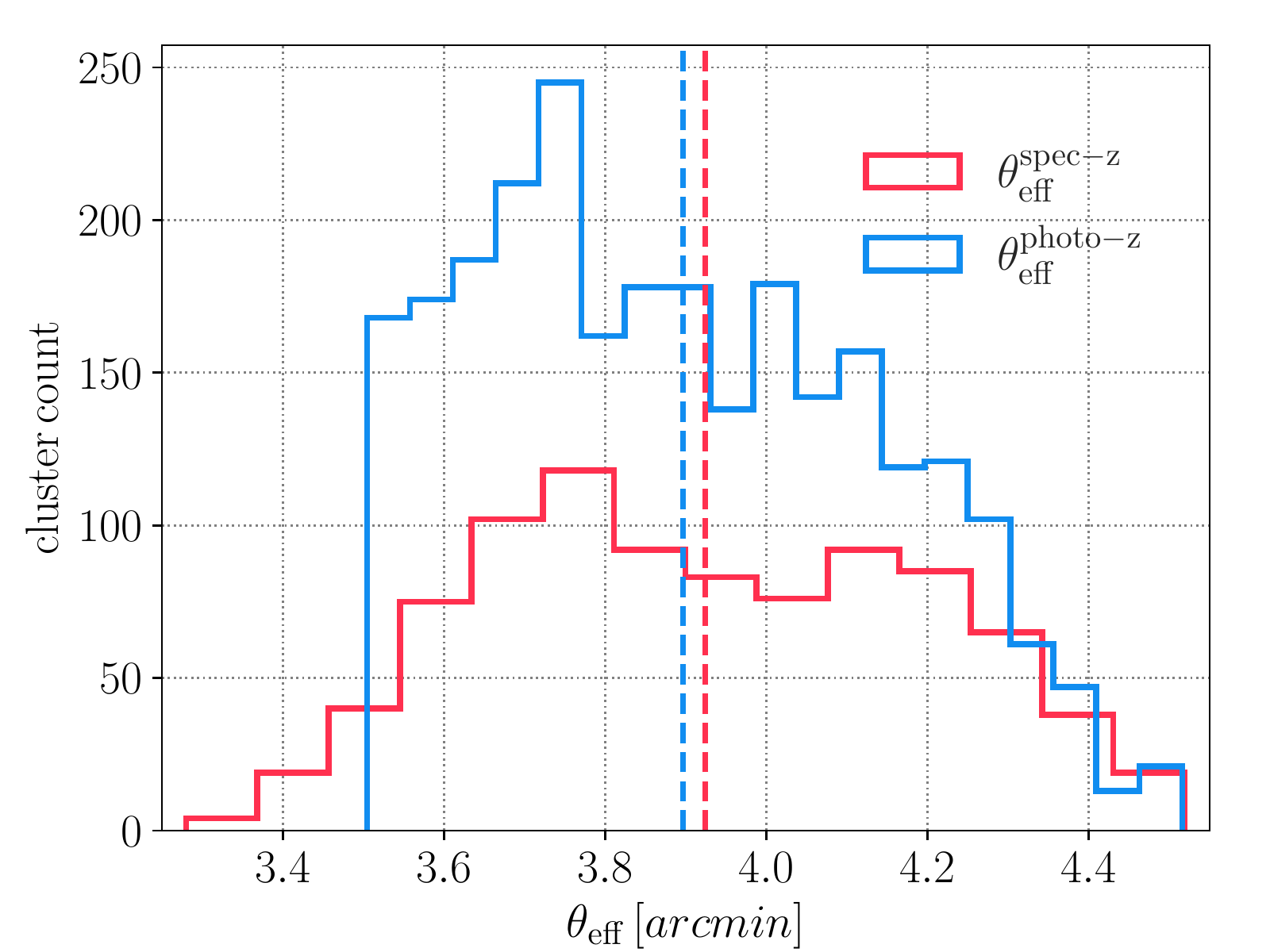}
        }}
    \caption{Redshift (left) and apparent size (right) distributions of 3512 maxBCG clusters selected for our analysis. The vertical lines present the mean of redshift distributions for spec-z and photo-z set, respectively.}\label{fig:maxBCG_histograms}
\end{figure*}

The mean and median mass of our cluster sample (including both datasets) are $M^{\mathrm{mean}}_{200}=7.18\times10^{13}\Msun$ and $M^{\mathrm{median}}_{200}=7.31\times10^{13}\Msun$, while the mean and median redshift are $z^{\mathrm{mean}}=z^{\mathrm{median}}=0.257$. This results in a mean and a median apparent size of $\theta^{\mathrm{mean}}_{\mathrm{eff}}=\theta^{\mathrm{median}}_{\mathrm{eff}}=3.9'$.
The full histograms of redshift and apparent angular size for both sets of clusters are shown in the left and right panels of \reffig{maxBCG_histograms}, respectively.
The apparent size distributions shown in right panel of \reffig{maxBCG_histograms} are rather concentrated; their standard deviations are $\sim0.25$ arcmin. This actually implies that the AAP filter would not yield significant improvement, as compared to the traditional AP filter, for both sets of cluster considered here. We nevertheless adopt the former for the individual-scale measurements, as already noted in \refsec{kSZ:signal_extraction}, since it offers a more physical interpretation of the variation of the signal with scale. For the combined measurements in \refsec{kSZ:multi_scale_model}, \refsec{kSZ:multi_scale_AP_result}, and \refsec{kSZ:multi_scale_AP_test}, we adopt the latter as it simplifies the amount of computation involved.

\subsubsection{Modeling maxBCG cluster optical depth}
\label{sec:kSZ:tau_model}

\refeq{taudef} is written for a point source. If a cluster is resolved, we need to integrate \refeq{taudef} over the angular extent $\theta_{\eff,i}$ of the cluster. For example, assuming a spherically symmetric Gaussian profile for the electron gas, \refeq{taudef} becomes
\be
\tau(\theta_i, \theta) = \frac{\tau_{0,i}}{\sqrt{2\pi\theta_{\eff,i}^2}} \int_0^\theta \textrm{d}^2\theta\; \exp\left(-\frac{\left(\theta-\theta_i\right)^2}{2\, \theta_{\eff,i}^2}\right),
\label{eq:integrated_tau}
\ee
where $\tau_{0,i}$ is the \emph{integrated} optical depth specific for cluster $i$. We will return to this point below in the discussion following \refeq{datamodel}.

As the fraction of electrons held by the neutral gas is presumably small, we will assume that all baryons in all the clusters being considered here are fully ionized, i.e. setting $f_{\mathrm{free}}=1$ in $N_e=f_{\mathrm{free}}\fgas M_{200,i}/\mu_e m_p$. We further adopt a universal gas-mass fraction $\fgas=f_b\equiv\Omega_b/\Omega_m=0.16$ following the cosmological baryon abundance and a mean particle weight per electron $\mu_e=1.17$.
Our expression for the integrated cluster optical depth defined in \refeq{integrated_tau} then becomes
\be
   \tau_{0,i} = \frac{\s_T}{D^2_{A,i}}\frac{f_b M_{200,i}}{\mu_e m_p}.
   \label{eq:taucluster}
\ee


\subsection{\borg-SDSS3 reconstructed velocity field}
\label{sec:kSZ:velocity_data}

Below, we briefly summarize how the velocity field employed in our analysis is obtained through the process of Bayesian forward modeling, highlighting key features of the \borg{} algorithm and its output. We then detail how we model the large-scale bulk flow of maxBCG clusters using the reconstructed velocity field based on \borg-SDSS3 outputs.

\subsubsection{\borg-SDSS3 reconstruction}
We employ the \emph{non-linear} velocity field set by initial conditions within the SDSS3-BOSS survey volume. The latter is constrained in \cite{Lavaux:2019} using the Bayesian forward-modeling algorithm \borg{} \cite{Jasche:2013}. Taking the LOWZ and CMASS galaxies from the SDSS3-BOSS DR13 release \cite{Dawson:2013, Alam:2015} as input, \borg{} systematically explores a $256^3$-dimensional parameter space of the three-dimensional initial conditions -- augmented by the galaxy bias parameters. The algorithm does this essentially by following these three steps repeatedly:
\begin{enumerate}
	\item Forward-evolving each realization of a \emph{Gaussian} initial matter field from $z_{\mathrm{ini}}\sim1000$ to $z=0$. The amplitude of the initial fluctuations is set by the fixed, pre-chosen set of cosmological parameters. In \cite{Lavaux:2019}, the gravitational forward model used to evolve the initial fluctuations is the first-order Lagrangian perturbation theory \cite{Zeldovich:1970, Buchert:1992}.
	\item Apply a galaxy bias transformation, together with other observational effects, including redshift-space distortion, light-cone effect, survey selection function, foreground contamination, etc. (see Section 2.2-2.6 of \cite{Lavaux:2019} for further details), on the \emph{evolved} matter density field to achieve a \emph{predicted} galaxy field.
	\item Compare the \emph{predicted} and \emph{observed} galaxy field (here LOWZ and CMASS) to simultaneously constrain the initial and evolved matter density fields.
\end{enumerate}
The result is a fully probabilistic inference of the cosmic matter and velocity fields, taking into account known and unknown systematic effects (within some pre-defined limits \cite{Lavaux:2019}). The setup of the inference is given as follows.
The initial conditions are generated on a comoving grid consisting of $256^3$ cells and covering a comoving volume of $4000^3\,h^{-3}\;\mathrm{Mpc}^3$, which corresponds to a grid resolution of $L_{\mathrm{grid}}=15.624\Mpch$. The SDSS3-BOSS data are projected in a sub-volume with the observer located at $\vx=\{200,0,-1700\}\Mpch$ with respect to the center.
A total number of 10360 MCMC samples was collected \cite{Lavaux:2019}.
Initial power-spectrum analysis in \cite{Lavaux:2019} showed that the MCMC chain converged after $\sim2000$ samples, i.e. the chain has passed its burn-in phase and reached a stationary distribution after the first $\sim2000$ samples. Here, we consistently remove all samples whose identifier is less than $s=2000$. To facilitate the storage and processing of these samples, the chain is thinned by a factor of 10, i.e. we only include 1-in-every-10 samples in our analysis; more details can be found in \cite{Lavaux:2019}.

Taking initial conditions constrained by the \borg-SDSS3 reconstruction \cite{Lavaux:2019} as input, we run DM-only simulations with the same cosmological parameters adopted by the reconstruction using the \emph{temporal} COmoving Lagrangian Acceleration algorithm \cite{Tassev:2013} (\tcola{} hereafter), and the cloud-in-cell (CIC) projection of particles, to obtain the large-scale velocity field at the maxBCG catalog mean redshift $z=0.23$.
This includes in total 837 \tcola{} simulations, one for each of our constrained initial conditions, at the resolution of $N_{\mathrm{part}}=1024^3$. These are used for the estimation of cluster LOS velocities as well as their uncertainties.

We additionally generate a GADGET-2 \cite{Springel:2005} simulation at resolution of $N_{\mathrm{part}}=2048^3$ from initial conditions specified by the sample $s=9000$ of \borg-SDSS3 reconstruction (see \refapp{kSZ:mock} for details). We use this full N-body, high resolution simulation to specifically:
\begin{enumerate}
   \item estimate the small-scale motion of clusters unresolved by the \borg-SDSS3 reconstruction and \tcola{} re-simulation (see \refsec{kSZ:bulkflowmodel}),
   \item verify that our kSZ estimators are unbiased (see \refsec{kSZ:individual_scale_model}),
   \item measure the cluster signal profile (see \refsec{kSZ:multi_scale_model}).
\end{enumerate}

\subsubsection{Modeling the large-scale bulk flows of galaxy clusters}
\label{sec:kSZ:bulkflowmodel}
\begin{figure}
    \centerline{\resizebox{0.7\textwidth}{!}{\includegraphics*{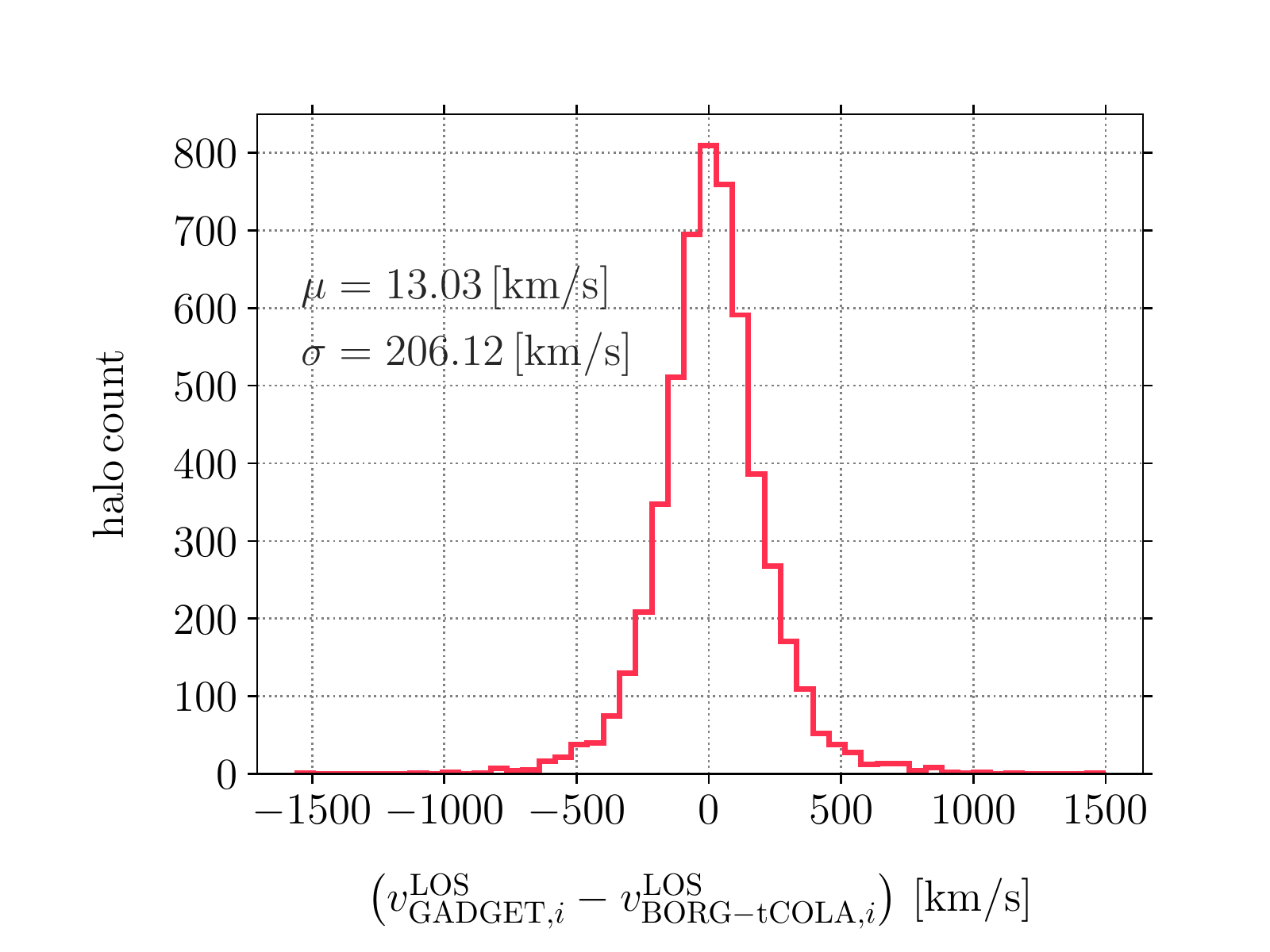}
    }}
    \caption{Histogram of $\left(v^{\LOS}_{\textsc{nbody}, i} - v^{\LOS}_{\tcola, i}\right)$ measured from DM halos within LOWZ-CMASS volume in our GADGET-2 and \tcola{} simulations of the \borg-SDSS3 sample $s=9000$. We apply the same redshift and mass cuts that are applied for maxBCG clusters. The vertical line represents the sample mean. The sample variance is measured at $\sigma^2=4.67\times10^4\,\mathrm{km}^2 \mathrm{s}^{-2}$.}\label{fig:sigma2_vlos}
  \end{figure}

We model the large-scale bulk flow of galaxy clusters in \refeq{Tksztau} as a sum of two components:
\be
v^{\LOS}(\vx, \vnhat) = v^{\LOS}_L(\vx, \vnhat) + \eps^{\LOS}_S(\vx, \vnhat)\,,
\label{eq:vcluster}
\ee
where $v^{\LOS}_L$ is the \emph{large-scale} LOS bulk-flow estimated from the \borg-SDSS3 reconstruction posterior while $\eps^{\LOS}_S$ is the unresolved \emph{small-scale} LOS velocity.
We further assume that, for all clusters:
\be
\eps^{\LOS}_S \sim \N\left(0, \sigma_{\eps^{\LOS}_S}\right)\,,
\label{eq:vnoise}
\ee
where $\sigma_{\eps^{\LOS}_S}\simeq206.12\,\mathrm{km} \mathrm{s}^{-1}$ is estimated from the standard deviation of $\left(v^{\LOS}_{\textsc{nbody}, i} - v^{\LOS}_{\tcola, i}\right)$ distribution (see \reffig{sigma2_vlos}) in which $v^{\LOS}_{\textsc{nbody}, i}$ and $v^{\LOS}_{\tcola, i}$ refer to the LOS velocity of halo $i$ as respectively measured from the previously mentioned GADGET-2 simulation and from \tcola{} simulation of the same \borg-SDSS3 sample, $s=9000$.


\section{Data model and kSZ posterior}
\label{sec:kSZ:datamodel}

Given an inferred ensemble of the large-scale LOS velocity of each galaxy cluster $i$, $v^{\LOS}_{L,i}$, our goal is to construct a likelihood for the extracted kSZ signals at locations of all galaxy clusters in our sample. Below, we will derive this likelihood in two cases of input data:
\begin{enumerate}
   \item The single-cluster signal is extracted at individual physical scales. This yields multiple measurements of the signal, each using information from a specific scale.
   \item The single-cluster signal is simultaneously extracted at multiple scales. This yields one single measurement of the signal combining information from all scales.
\end{enumerate}
While the former can be applied for an analysis focusing on the study of cluster gas profile, we expect the latter to be a more sensitive measurement, as information from all scales is combined and the impact of CMB noise on large scales can be reduced. Our derivations assume that the kSZ measurements at individual cluster locations are independent, i.e. there is no significant overlap between the AP/AAP filters. We verified this is indeed the case for our cluster sample selected from the maxBCG catalog (see \refsec{kSZ:maxbcg_catalog}).


\subsection{kSZ likelihood: single angular scale}
\label{sec:kSZ:individual_scale_model}

Let us express our data model as
\be
\D T_{\kSZ, i}^{\theta_f}/T_0 = -\alpha^{\theta_f} \, \tau_i\, v^{\LOS}_{L,i}/c - \tau_i\, \eps^{\LOS}_{S, \it i}/c + \eps^{\theta_f}_{0, i} \, ,
\label{eq:datamodel}
\ee
where we have introduced $\alpha^{\theta_f}$ as the amplitude of the kSZ signal from the \emph{large-scale} bulk flow of cluster $i$. Here and below, we simply adopt $\tau_i=\tau_{0,i}$. For cases where the filter size is smaller than that of the cluster, this means letting $\alpha^{\theta_f}$ absorb all specific details about the cluster gas profile.

We expect to measure a value of $\alpha^{\theta_f}$ consistent with zero in the case of no detection, whereas a value of order of unity at filter sizes that are large enough to encompass the whole cluster, i.e. $\theta_f \geq \sqrt{\theta^2_{\mathrm{vir}} + \theta^2_{\beam}}$, corresponds to the expectation from the simple model of optical depth discussed in \refsec{kSZ:tau_model}.
The exact value of $\alpha$ depends on that of $f_{\mathrm{free}}$ (cf. \refeq{taucluster}) and on the amount of ionized gas outside but associated with the cluster. Further, it is also sensitive to systematics in the \emph{amplitude} of the reconstructed velocity field and in the weak lensing mass calibration of galaxy clusters \cite{Rozo:2009}. Hence a direct interpretation of $\alpha$ as $\fgas$ would require careful modeling of these uncertainties. In this paper, we focus on the significance of the kSZ signal detection, which is fortunately not significantly affected by any of those.

In \refeq{datamodel}, the first noise term on the r.h.s is induced by unresolved small-scale LOS velocity $\eps^{\LOS, s}_{S, \it i}$ (see \refsec{kSZ:bulkflowmodel}). Note that this term scales with cluster optical depth $\tau_i$. It is thus typically negligible for the clusters considered in our analysis.
The second noise term $\eps^{\theta_f}_{0, i}$ denotes the residual of primary CMB anisotropies plus inhomogeneous instrumental noise, which we assume to be a Gaussian random noise with zero mean and variance $\left(\sigma^{\theta_f}_{0, i}\right)^2$.

\refeq{datamodel} holds so long as the tSZ plus other foreground contaminations cancel out. A significant degree of cancelation is expected since they are, to first order, uncorrelated with the LOS large-scale velocity $v^{\LOS}_{L,i}$.
Given the cluster mass cut introduced in \refsec{kSZ:maxbcg_catalog}, we confirmed that this condition holds (see \refapp{kSZ:masscut}).

Both the signal amplitude and noise are functions of the AP filter size $\theta_f$ (or AAP filter scale $\varphi_f$) -- as indicated by the superscript $\theta_f$. However, for the sake of readability, we will omit the superscript $\theta_f$ in all following equations.

The corresponding likelihood distribution for a single velocity realization -- as inferred by a \borg-SDSS3 sample $s$ -- is given by
\begin{multline}
   \P\left(\{\D T_{\kSZ, i}/T_0\} \Big|\alpha,\, \{\tau_i v^{\LOS, s}_{L,\it i}/c\} \right)  =  \\
   \prod_i \frac{1}{\sqrt{2\,\pi \, \sigma_i^2}} \exp\left\{-\frac{1}{2\sigma^2_i} \, \left (\frac{\D T_{\kSZ, i}}{T_0} +  \alpha \, \tau_i \, \frac{v^{\LOS, s}_{L, i}}{c} \right)^2\right\}
   \label{eq:likelihood_singlesample}
\end{multline}
where
\be
\sigma^2_i \equiv \sigma^2_{0, i} + (\tau_i/c)^2\sigma^2_{\eps^{\LOS}_S}.
\label{eq:sigma2}
\ee

We now seek to construct a posterior distribution for $\alpha$, marginalized over $N$ velocity realizations, i.e.
\bab
\P\left(\alpha \Big|\{\D T_{\kSZ, i}/T_0\}\right)  &= \int \mathrm{d}\{x_i\}\, \P\left(\alpha,\, \{x_i\} \Big|\D T_{\kSZ, i}/T_0 \right) \\
&\propto  \P\left(\alpha\right)\, \int \mathrm{d}\{x_i\} \, \P\left(\{x_i\}\right) \P\left( \D T_{\kSZ, i}/T_0 \Big|\alpha,\, \{x_i\} \right) \,,
\label{eq:posterior_marginalization}
\eab
where we have used $\{x_i\}\equiv\{\tau_i \, v_{\LOS, i}/c\}$ and introduced the prior on $\alpha$ explicitly as $\P(\alpha)$.
The \borg-SDSS3 reconstruction provides a sampled approximation
\be
\P\left( \{x_i\} \right) \approx \frac{1}{N} \sum_{s=1}^N \delta^D\left(\{x_i\}-\{x_i\}^s\right) \, ,
\ee
where $\delta^D(\{x_i\})$ denotes the Dirac delta distribution.
We thus can rewrite \refeq{posterior_marginalization} as
\ba
\P\left(\alpha \Big|\{\D T_{\kSZ, i}/T_0\}\right)
&\propto  \P\left(\alpha\right) \frac{1}{N} \sum_{s=1}^N  \prod_i \frac{1}{\sqrt{2\,\pi \, \sigma_i^2}} \exp\left\{-\frac{1}{2\sigma^2_i} \, \left (\frac{\D T_{\kSZ, i}}{T_0} +  \alpha \, x^s_i \right)^2\right\} \, , \\
&\propto \P\left(\alpha\right)  \sum_s^N \lambda_s \frac{\exp\left(-\,\frac{\left(\alpha-\mu_s\right)^2}{\,2\left(\sigma_s\right)^2}\right)}{\sqrt{2\pi\,\left(\sigma_s\right)^2 }}\, ,
\label{eq:marginal_posterior}
\ea
which is a mixture of Gaussian distributions. Each component of the mixture consists of an individual velocity realization -- indexed by $s$ -- associated with a mixture weight $\lambda_s$, which is given by (see \refapp{kSZ:gaussianmixture} for a detailed derivation)
\be
\lambda_s = \frac{\exp\left[\omega_s \, + \frac{1}{2}\mathrm{ln}(2\pi\,\left(\sigma_s\right)^2 )\right]}{ \sum_s^N \exp\left[\omega_s \, + \frac{1}{2}\mathrm{ln}(2\pi\,\left(\sigma_s\right)^2 )\right]},
\label{eq:lambda_s}
\ee
in which
\be
\sigma_s^2 = \left[ \sum_i \left( \frac{x^s_i}{\sigma_i}\right)^2 \right]^{-1},
\label{eq:sigma2_s}
\ee
and
\be
\omega_s \equiv \mu^2_s / \left(2\sigma^2_s\right),\,\,\,\, \mu_s = \frac{\sum_i \left[ \left(\D T_{\kSZ, i}/T_0\right)\,x^s_i\right] / \sigma_i^2}{\sum_i \left(x^s_i/\sigma_i\right)^2}.
\label{eq:omega_s}
\ee
Note that the weights $\lambda_s$ give preference to better-fitting realizations of the peculiar velocity field, and $\sum_s^N \lambda_s=1$.

For simplicity, in what follows, we assume a \emph{uniform} prior on $\alpha$ such that $\P(\alpha)=1$ for all \borg-SDSS3 velocity realizations. The posterior mean estimate of $\alpha$ is then given by (see \refapp{kSZ:gaussianmixture} for details)
\be
\<\alpha \>_s = \sum_s^N \lambda_s\mu_s = \sum_s^N \lambda_s \,\frac{\sum_i \left[\left(\D T_{\kSZ, i}/T_0\right)\,x^s_i\right]/\sigma_i^2}{\sum_i \left(x^s_i/\sigma_i\right)^2},
\label{eq:alpha_ensemble_mean}
\ee
while its variance is given by
\be
\sigma^2_{\alpha} = \sum_s^N \lambda_s  \sigma_s^2  + \sum_s^N \lambda_s \,\left( \mu_s-\langle \alpha \rangle_s\right)^2.
\label{eq:alpha_ensemble_variance}
\ee
In the Bayesian language, the significance of each kSZ measurement in supporting the positive-kSZ hypothesis (against the no-kSZ hypothesis) would be quantified through the Bayes factor. In our particular case, this factor can be directly computed as the ratio between two cumulative distribution functions (CDF) of positive-kSZ $\P(\alpha>0)$ and no-kSZ hypothesis $\P(\alpha\leq0)$:
\be
BF = \frac{\P\left(\alpha>0\Big|\{\D T_{\kSZ, i}/T_0\}\right)}{\P\left(\alpha\leq0\Big|\{\D T_{\kSZ, i}/T_0\}\right)}.
\label{eq:bayes_factor}
\ee
Previous kSZ measurements in literature, however, usually reported their significance in terms of distance between the measurement and the value of the null hypothesis, measured in multiples of standard deviation $\sigma$ of a normal distribution \citep[see, e.g.][]{Planck:2016a, Hernandez:2015, Schaan:2016, Lim:2017, Tanimura:2020}. To facilitate a direct comparison, we adopt the same unit by matching the CDF of the normal distribution to that of our posterior distribution, i.e.
\be
\P_{\mathrm{normal}}\left(x\leq-\mathrm{sign.}\right) = \P(\alpha\leq0) \,,
\label{eq:significance}
\ee
and sign. would be our significance measured in the familiar unit of $\sigma$.

\begin{figure}
    \centerline{\resizebox{0.7\textwidth}{!}{\includegraphics*{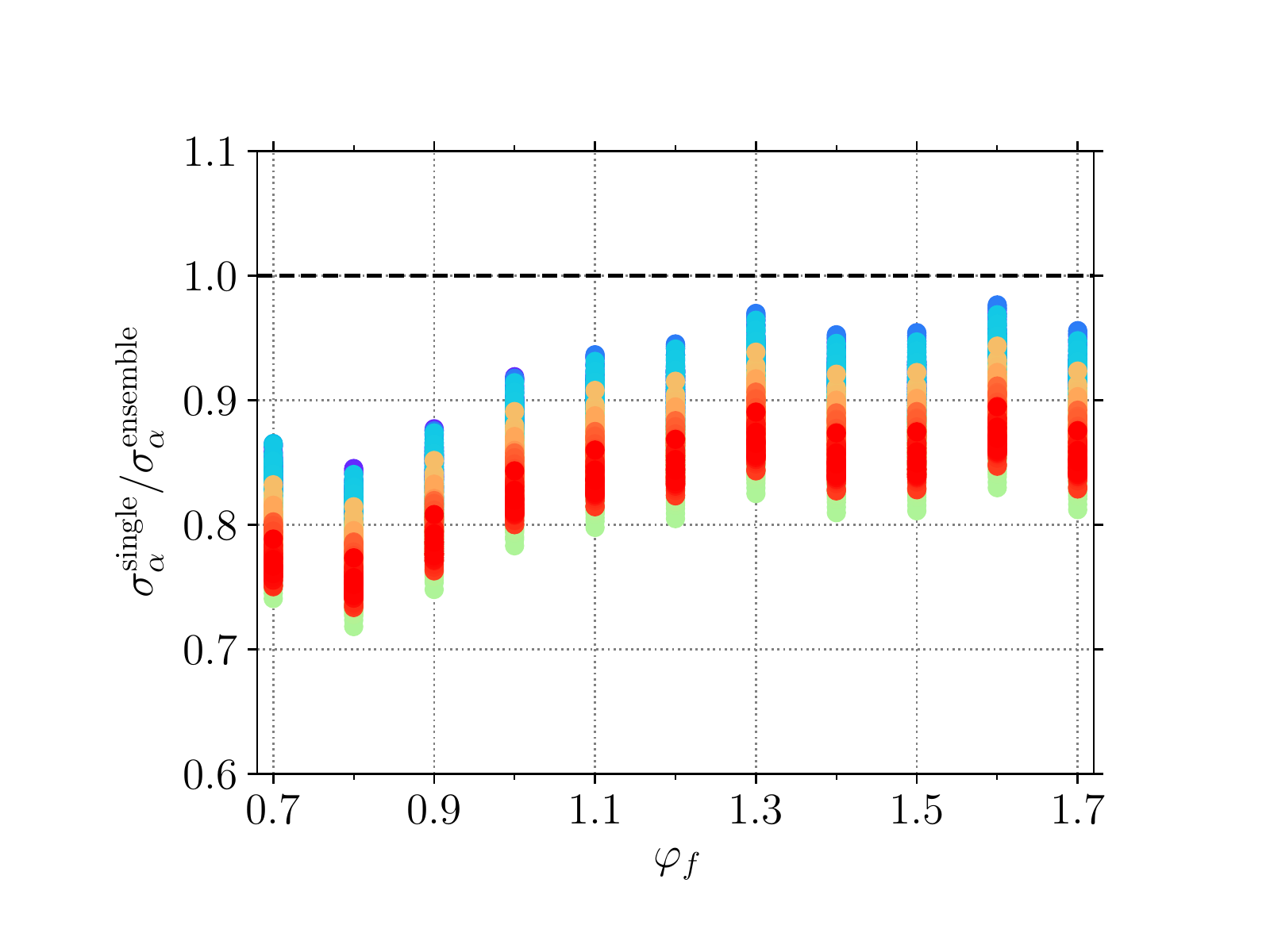}
    }}
    \caption{Ratio between the uncertainty on $\<\alpha\>_s$ when having only one single realization of the peculiar velocity field, as in the cases reported so far in the literature, and that when having instead an ensemble of realizations, as in our case. Each color point represents a realization of the reconstructed velocity field, corresponding to a \borg-SDSS3 MCMC sample.}\label{fig:sigma_single_vs_ensemble_VLOS}
  \end{figure}

Note that \refeq{alpha_ensemble_mean} is identical to Eq. (9) in \cite{Schaan:2016} if one takes all $\lambda_s=1$ and neglects the uncertainty in the velocity reconstruction process.
As can be seen in \refeq{alpha_ensemble_variance}, our estimator properly includes systematic uncertainties in this step. To better illustrate this point, in \reffig{sigma_single_vs_ensemble_VLOS}, we compare the uncertainty on the inferred amplitude when one only has a single velocity realization $s$, and when one has an ensemble of realizations $\{s\}$.
The single realization case consistently \emph{underestimates} the uncertainty by -- from small to large scales -- $\simeq10-20\%$.
It is worth pointing out also that different realizations yield different estimates of $\alpha_s$ itself. A combination of, say, high $\alpha_s$ and low $\sigma_{\alpha_s}$, can significantly bias the significance of any kSZ detection that does not account for uncertainties in velocity reconstruction. We note that this trend might explain the large variation in detection significance of previous kSZ measurements, as, for example, summarized in Tab.~1 of \cite{Planck:2018LIII}. In our case, the significance can be biased by as high as a factor of 2 if one cherry-picks only one single best-fitting \borg-SDSS3 velocity realization.

We further test our estimator in \refeq{alpha_ensemble_mean} on mock input data where a kSZ signal template -- generated by DM halos in the GADGET-2 simulation of the \borg-SDSS3 sample $s=9000$ (see \refapp{kSZ:mock}), assuming a Gaussian gas profile -- is injected into a \smica2018-like CMB map (including both primary CMB and instrumental noise). By artificially varying the noise level, we verified that our estimator is indeed \emph{unbiased}. In the limit of vanishing noise, we obtain $\lambda_{s=9000}\to1$ while the other weights go to zero, correctly singling out the original \borg-SDSS3 sample used to generate the mock signal template.


\subsection{kSZ likelihood: combined signal}
\label{sec:kSZ:multi_scale_model}

\begin{figure}
    \centerline{\resizebox{0.7\textwidth}{!}{\includegraphics*{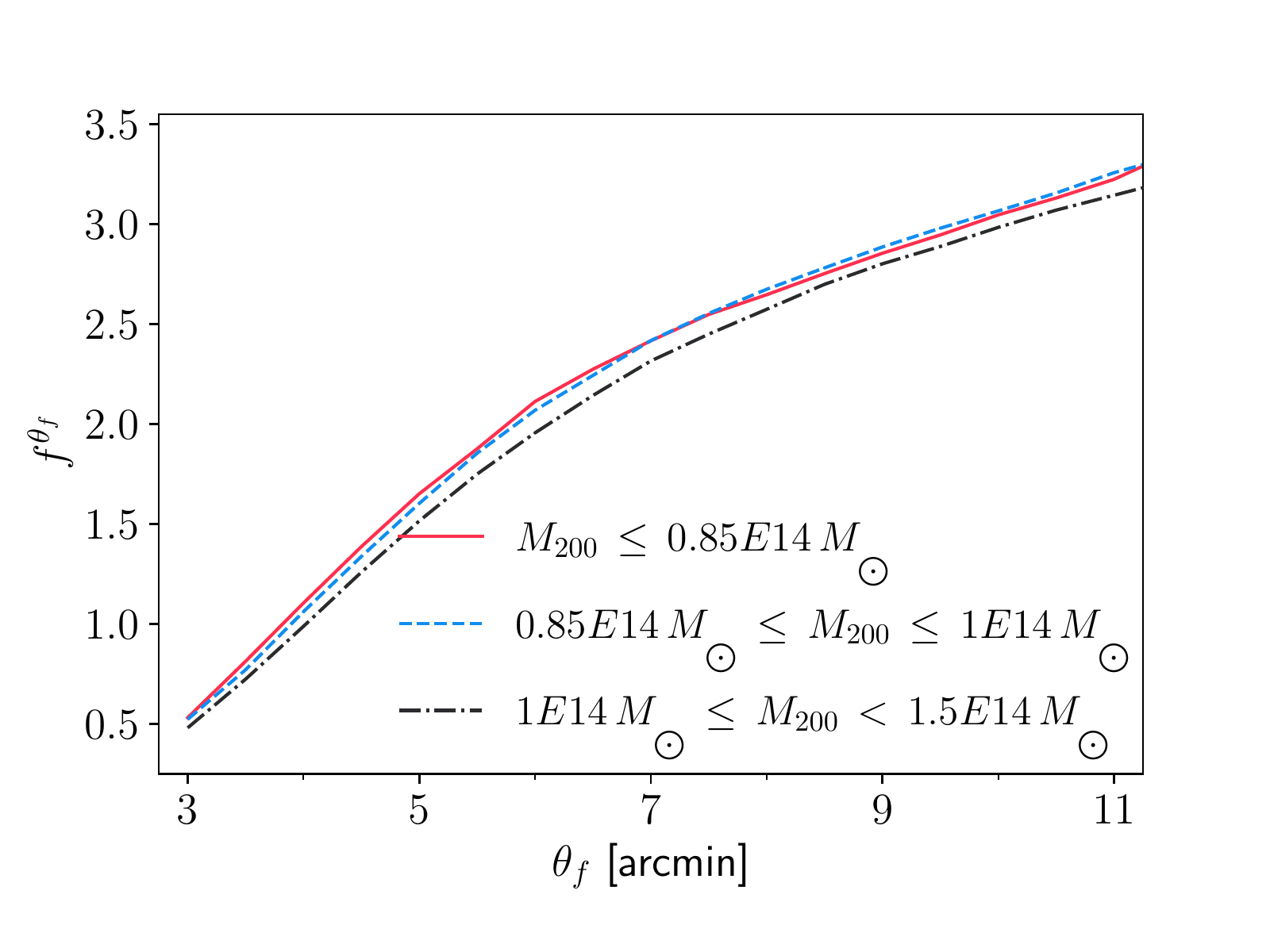}
    }}
    \caption{The profile $f^{\theta_f}$ at locations of DM halos measured from our mock kSZ template generated by member particles in the GADGET-2 simulation detailed in \refapp{kSZ:mock}. As a consistency check, we divide our DM halo sample of $M_{200}\leq1.5\times10^{14}\Msun$ into three mass bins as shown in the plot. Below $M_{200}\leq10^{14}\Msun$, the profile is not very sensitive to $M_{200}$. Only the result for $M_{200}\leq0.85\times10^{14}\Msun$ is used in our analysis.}\label{fig:gas_profile_measurement}
  \end{figure}

To combine measurements at different filter sizes $\theta_f$, it is necessary to modify \refeqs{datamodel}{marginal_posterior} to include the cluster gas profile $f(\theta_f)$ as
\be
\tau_i = \tau_{i,0}\,f^{\theta_f},
\label{eq:introduce_profile}
\ee
where we have assumed that this profile is universal. In this work, we obtain an estimate this profile by applying our pipeline on a pure-kSZ signal template generated from individual DM particles found in the GADGET-2 high-resolution simulation (see details in \refapp{kSZ:mock}). We show below, in \reffig{gas_profile_measurement}, the measurement of $f^{\theta_f}$ at locations of corresponding DM halos identified by Rockstar halo finder\footnote{\url{https://bitbucket.org/gfcstanford/rockstar/}} \cite{Behroozi:2013, Knebe:2013}, an adaptive hierarchical friends-of-friends (FoF) algorithm in six-dimensional phase-space. Note that we use the same velocity field to assign LOS velocity to the DM particles and halos. For simplicity, in case of the combined signal, we restrict ourselves to the AP filter whose $\theta_f$ does not depend on the cluster effective apparent size.

\refeq{datamodel} now becomes
\be
\vge{\D T}_{\kSZ, i}/T_0 = - \alpha \, \tau_{i,0} \vg{f} \, \left(v^{\LOS}_{L,i}/c\right) - \tau_{i,0} \vg{f} \, \left(\eps^{\LOS, s}_{S, \it i}/c\right) + \vg{\eps}_{0, i} \, ,
\label{eq:datamodel_combined}
\ee
where we have again omitted the superscript $\theta_f$ and instead expressed quantities that depend on $\theta_f$ as vectors to stress the fact that all measurements at different filter scales are now combined in \refeq{datamodel_combined}.
Note also that $\alpha$ is now simply a scalar instead of a function of $\theta_f$.
\refeq{likelihood_singlesample} can then be rewritten as
\begin{multline}
   \P\left(\{\vge{\D T}_{\kSZ, i}/T_0\} |\alpha,\, \tau_{i,0} \vg{f} \left(v^{\LOS, s}_{L, i}/c\right)\} \right)  \propto \prod_i |\C_i|^{-1/2} \\
   \quad \times \:\, \exp\left\{-\frac{1}{2} \, \left[\vge{\D T}_{\kSZ, i}/T_0 +  \alpha \, \tau_{i,0} \vg{f} \, \left(v^{\LOS, s}_{L, i}/c\right) \right]^\intercal \,\left(\C_i\right)^{-1}\, \left[\vge{\D T}_{\kSZ, i}/T_0 +  \alpha \, \tau_{i,0} \vg{f} \, \left(v^{\LOS, s}_{L, i}/c\right) \right]\right\}
\end{multline}
where the covariance matrix $\C_i$ for cluster $i$ is given by
\be
C^{\theta_f\theta_f'}_i=\left\langle\frac{\D T^{\theta_f}_{\CMB}(\v{\theta_i})}{T_0}\frac{\D T^{\theta_f'}_{\CMB}(\v{\theta_i})}{T_0}\right\rangle \, + \left\langle\frac{\D T^{\theta_f}_{\instr}(\v{\theta_i})}{T_0}\frac{\D T^{\theta_f'}_{\instr}(\v{\theta_i})}{T_0}\right\rangle \, + \, \tau^2_{i,0} f^{\theta_f} f^{\theta_f'} \sigma^2_{\eps^{\LOS}_S}\,,
\label{eq:covariance_matrix}
\ee
in which we have separated the primary CMB anisotropy and the instrumental noise into the first and second term, respectively.
Although, for completeness, we do include the last term on the r.h.s. of \refeq{covariance_matrix} in our covariance estimate, we note that it is negligible compared to the first two terms -- as it scales with $\tau^2$ -- and the exclusion of this term would affect neither the signal amplitude nor significance.
The second noise contribution is highly inhomogeneous due to the scanning strategy of the Planck satellite and requires instrument-specific mocks to estimate \cite{Planck:2015VIII, Planck:2018III}. To this end, it is worth mentioning that Planck does provide a limited set of 300 noise and residual systematics simulations for \smica2018 \cite{Planck:2018IV}. For this work, we instead choose to generate 2500 instrumental noise maps in which the noise value of each pixel is drawn from a zero-mean Gaussian whose variance is given by the corresponding temperature intensity variance in the Planck 2018 HFI Sky Map at frequency 143GHz \cite{Planck:2018III, Planck:2015VIII} (see details in \refapp{kSZ:AP_covariance}). While this estimate is likely conservative since \smica{} is a weighted linear combination of multiple frequency channels, we expect it to be robust and stable, especially at small filter sizes where this term dominates.
The first term could be estimated analytically using the Planck 2018 best-fit $\L$CDM power spectrum \cite{Planck:2018VI}. In the \emph{flat-sky} limit,
\be
\D T_{\CMB}^{\theta_f}(\v{\theta_i}) = \int \frac{d\vl}{(2\pi)^2} \, \exp{\left(i\vl\cdot\v{\theta_i}\right)} \, (\pi\theta_f^2) \, W(\l\theta_f) \, \D T_{\CMB}^{\obs}(\vl),
\label{eq:cmbap_Fourier}
\ee
where $W(l\theta_f)$ is the Fourier transform of the AP filter
\be
W\left(\l\theta_f\right) = 2\, \left[W_{\rm TH} \left(\l\theta_f\right)\, -\, W_{\rm TH}\left(\sqrt{2}\l\theta_f\right)\right],\qquad W_{\rm TH}\left(\l\theta_f\right) = 2\frac{J_1\left(\l\theta_f\right)}{\l\theta_f},
\ee
while $\vl$ is the two-dimensional wavevector perpendicular to the LOS, $\l=|\vl|$, and
\be
\D T_{\CMB}^{\obs}(\vl) = \D T_{\CMB}(\vl) B(\vl).
\label{eq:cmbobs}
\ee

\begin{figure}[htbp!]
   \centering
   \includegraphics[width=0.495\textwidth]{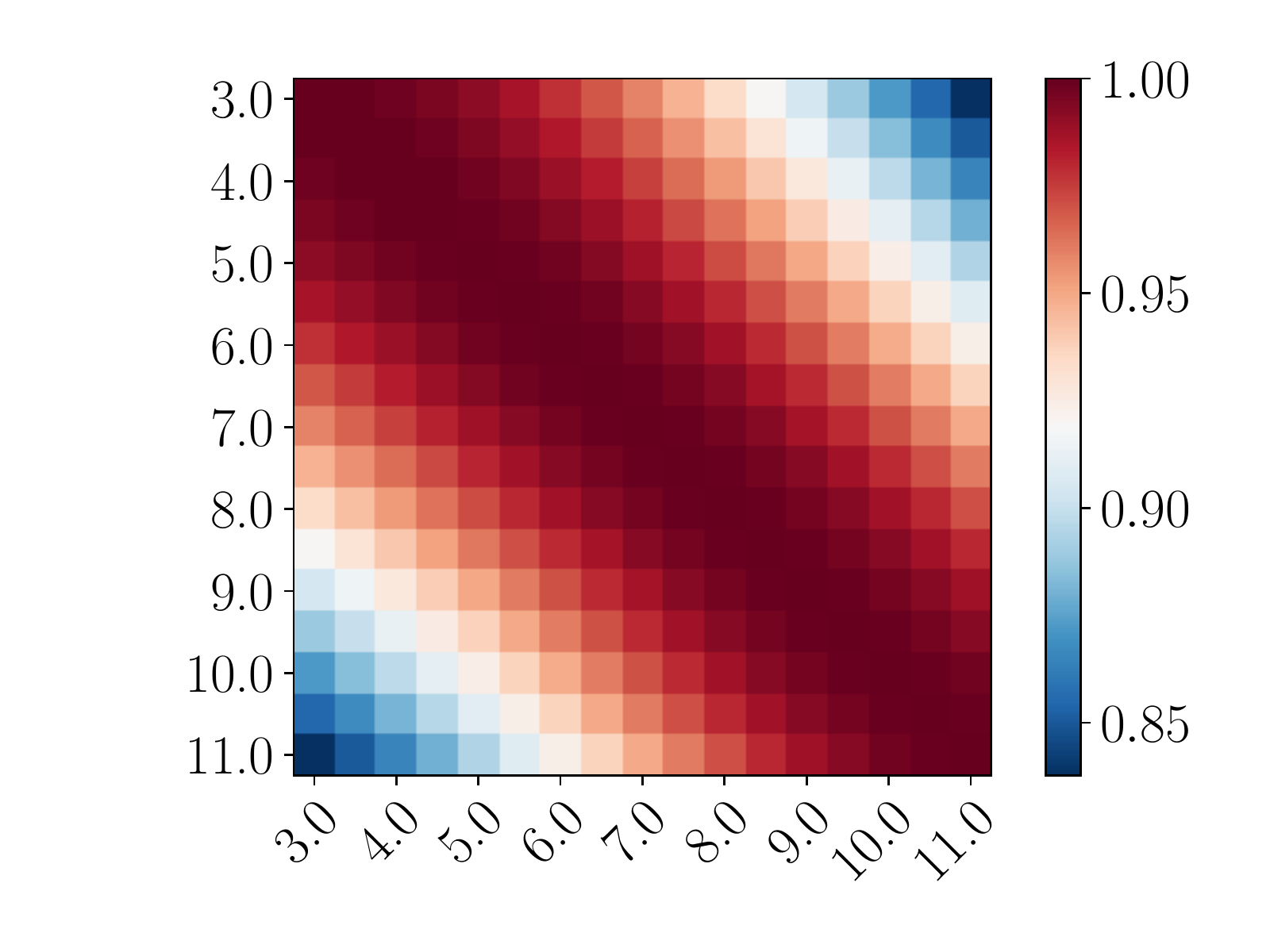}
   \includegraphics[width=0.495\textwidth]{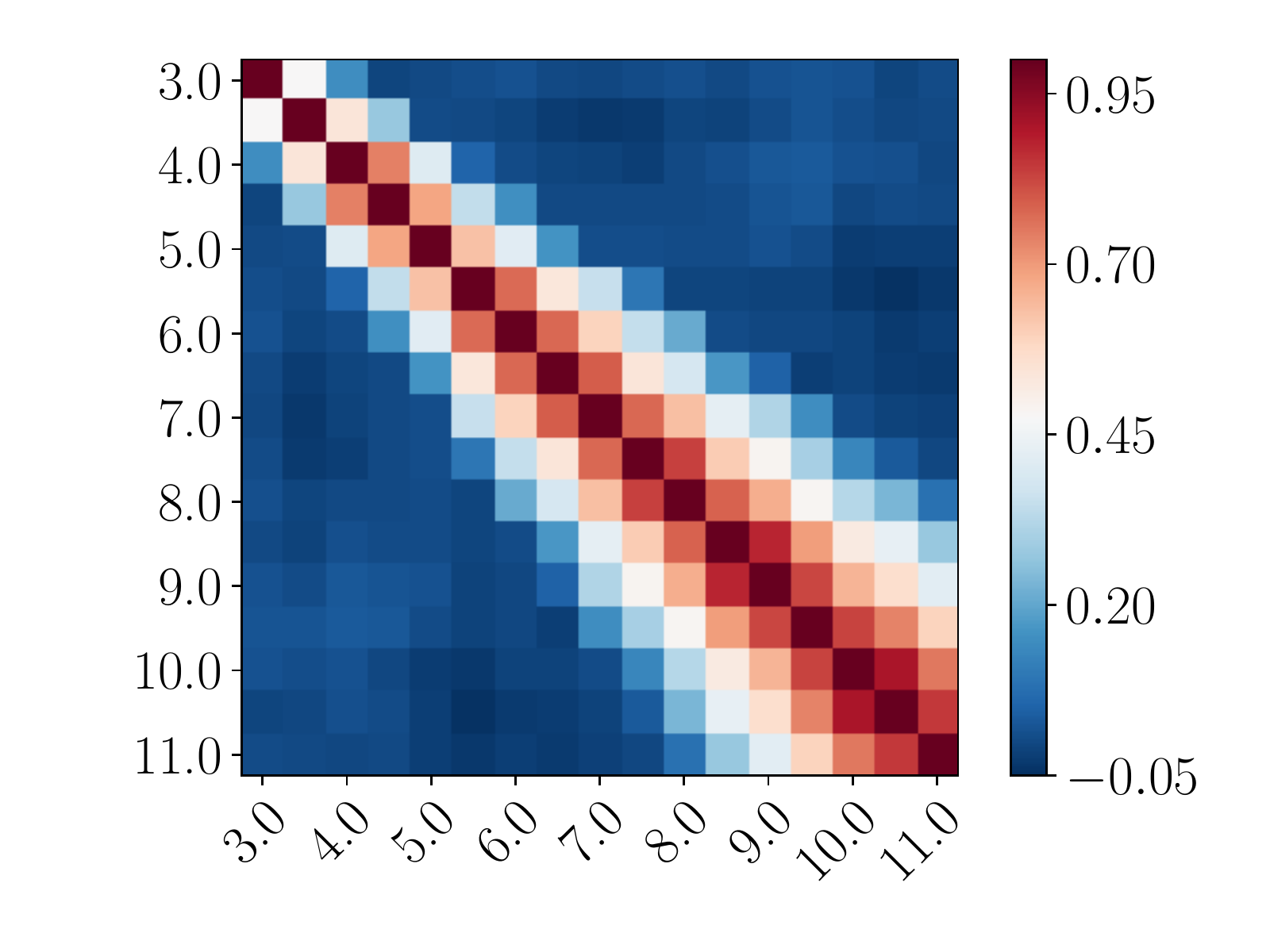}
   \includegraphics[width=0.495\textwidth]{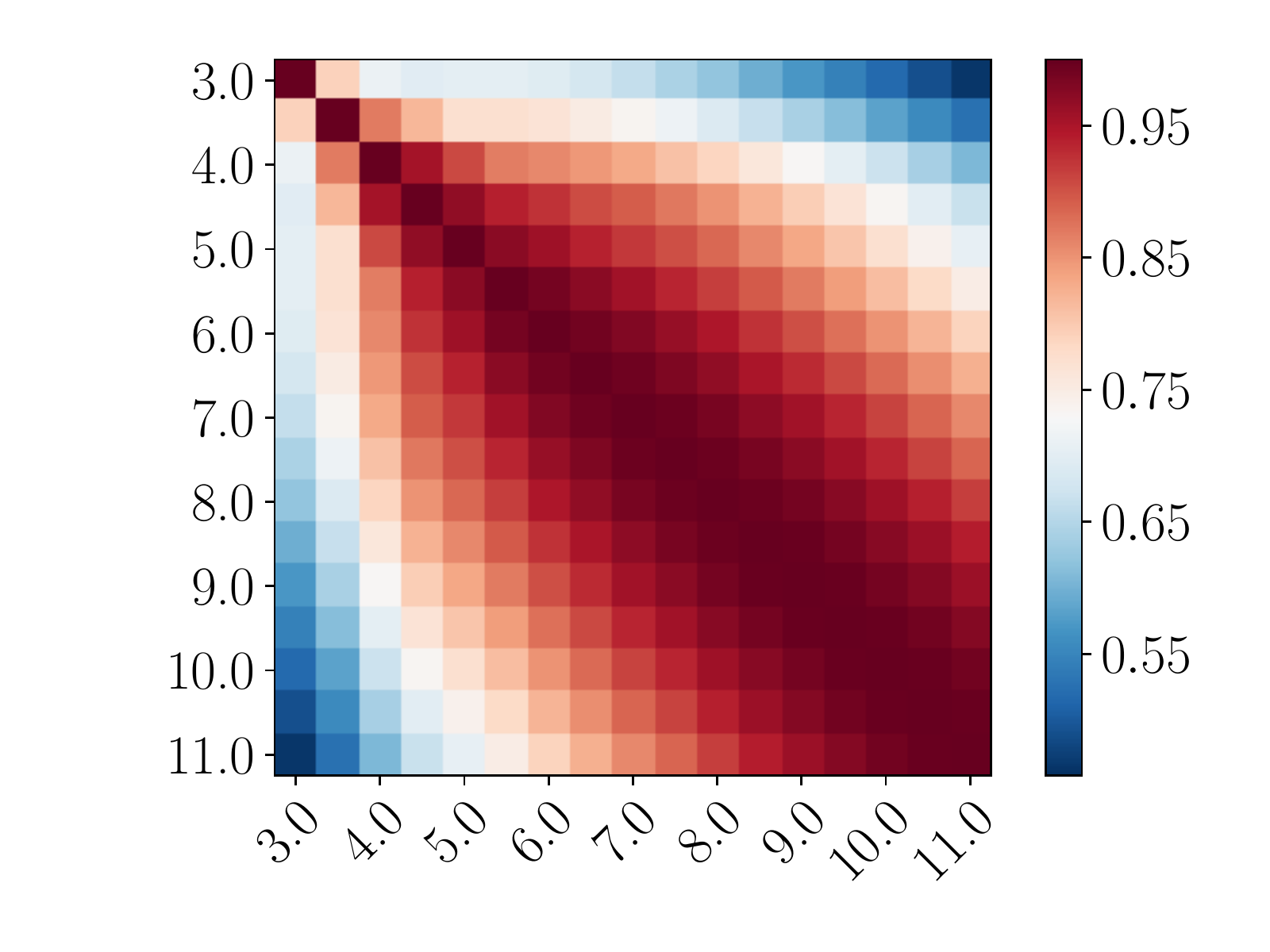}
   \caption{\textit{Top:} Correlation matrices of the primary CMB (left panel) and instrumental noise (right panel), as described by the first and second term on the r.h.s. of \refeq{covariance_matrix}.
      \textit{Bottom:} The total noise correlation matrix, evaluated using the full r.h.s. of \refeq{covariance_matrix}.
      \label{fig:Rij_AP}}
\end{figure}

The CMB covariance matrix in \refeq{covariance_matrix} is then explicitly given by (see \refapp{kSZ:AP_covariance})
\be
\C^{\theta_f\theta_f'}_{\CMB, i} = \frac{\pi\theta_f^2(\theta'_f)^2}{2T_0^2}\,\int_0^\infty d\l \, \l \, W\left(\l\theta_f\right)\, W\left(\l\theta_f'\right) \, \Cl{\CMB},
\label{eq:Fourier_covariance_matrix}
\ee
where $\Cl{\CMB}$ denotes the Planck 2018 $\L$CDM best fit angular power spectrum.
We show in \reffig{Rij_AP} the CMB correlation matrix evaluated from \refeq{Fourier_covariance_matrix}, the instrumental noise correlation matrix estimated, for one single cluster, from 2500 instrumental noise mocks, and the corresponding total correlation matrix of the AP measurements (including both primary CMB and instrumental noise contributions) at different filter sizes estimated by \refeq{covariance_matrix}. Note that the last two vary between cluster locations due to the inhomogeneity of the Planck instrumental noise.

We use the covariance matrix estimated by \refeq{covariance_matrix} for both the individual-scale and combined measurements of the signal. Specifically, for the individual-scale case which employs the AAP filter, we interpolate $\sigma^2_{\CMB, i}$ (cf. \refeq{sigma2}) from a 1024x1024 $\C^{\theta_f\theta_f'}_{\CMB}$ matrix.

The posterior of $\alpha$ given measurements at all filter sizes can be constructed similarly to \refeq{marginal_posterior}
\bab
\P\left(\alpha|\{\vge{\D T}_{\kSZ, i}/T_0\}\right)
&\propto  \P\left(\alpha\right) \frac{1}{N} \sum_{s=1}^N \prod_i  |\vg{C}_i|^{-1/2} \\
&\exp\left\{-\frac{1}{2} \, \left[\vge{\D T}_{\kSZ, i}/T_0 +  \alpha \, \vg{x}^s_i \right]^\intercal \,\left(\vg{C}_i\right)^{-1}\, \left[\vge{\D T}_{\kSZ, i}/T_0 +  \alpha \, \vg{x}^s \right]\right\}
\label{eq:combined_marginal_posterior}
\eab
where we have similarly used $\vg{x}^s_i \equiv \left\{\tau_{i,0} \vg{f} \, \left(v^{\LOS, s}_{L, i}/c\right)\right\}$.

The expressions of the posterior mean $\<\alpha\>_n$ and its variance $\sigma^2_\alpha$ are similar to those in \refeq{alpha_ensemble_mean} and \refeq{alpha_ensemble_variance} with modifications to $\mu_s$ and $\omega_s$ as described in \refapp{kSZ:gaussianmixture}.


\subsection{Modeling photo-z uncertainty}
\label{sec:kSZ:photoz_model}

\begin{figure}[htbp!]
   \centering
   \includegraphics[width=0.495\textwidth]{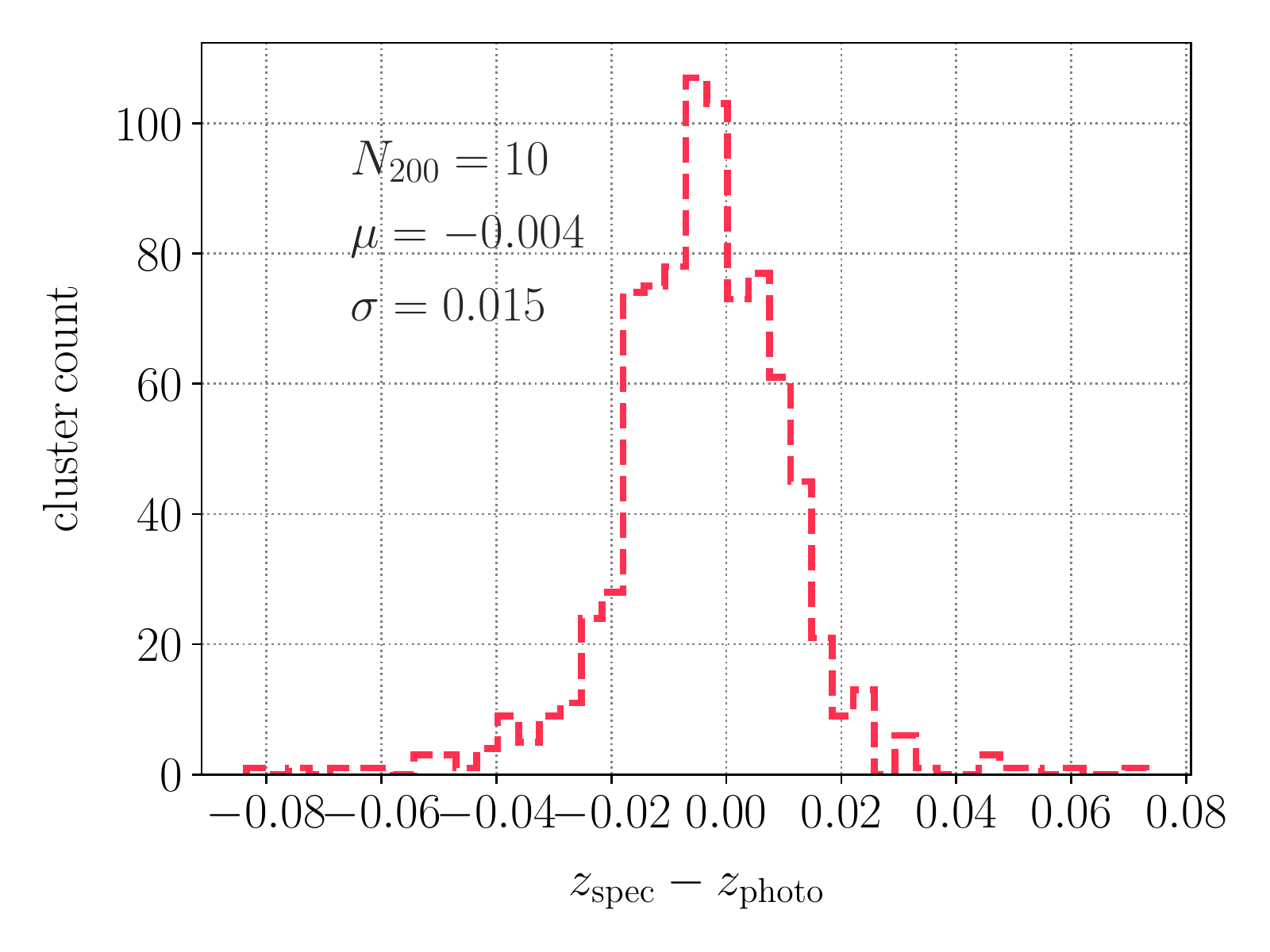}
   \includegraphics[width=0.495\textwidth]{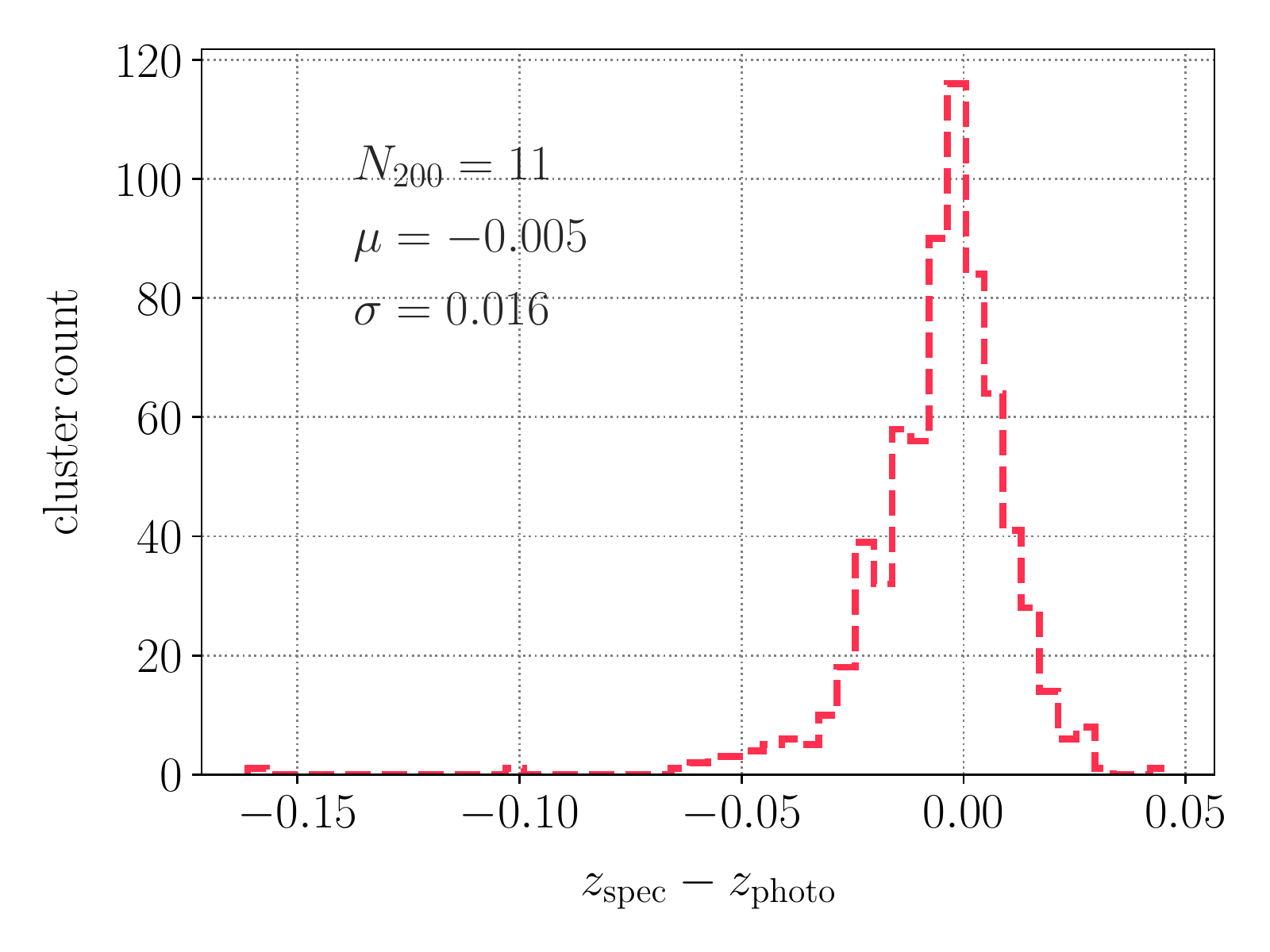}
   \includegraphics[width=0.495\textwidth]{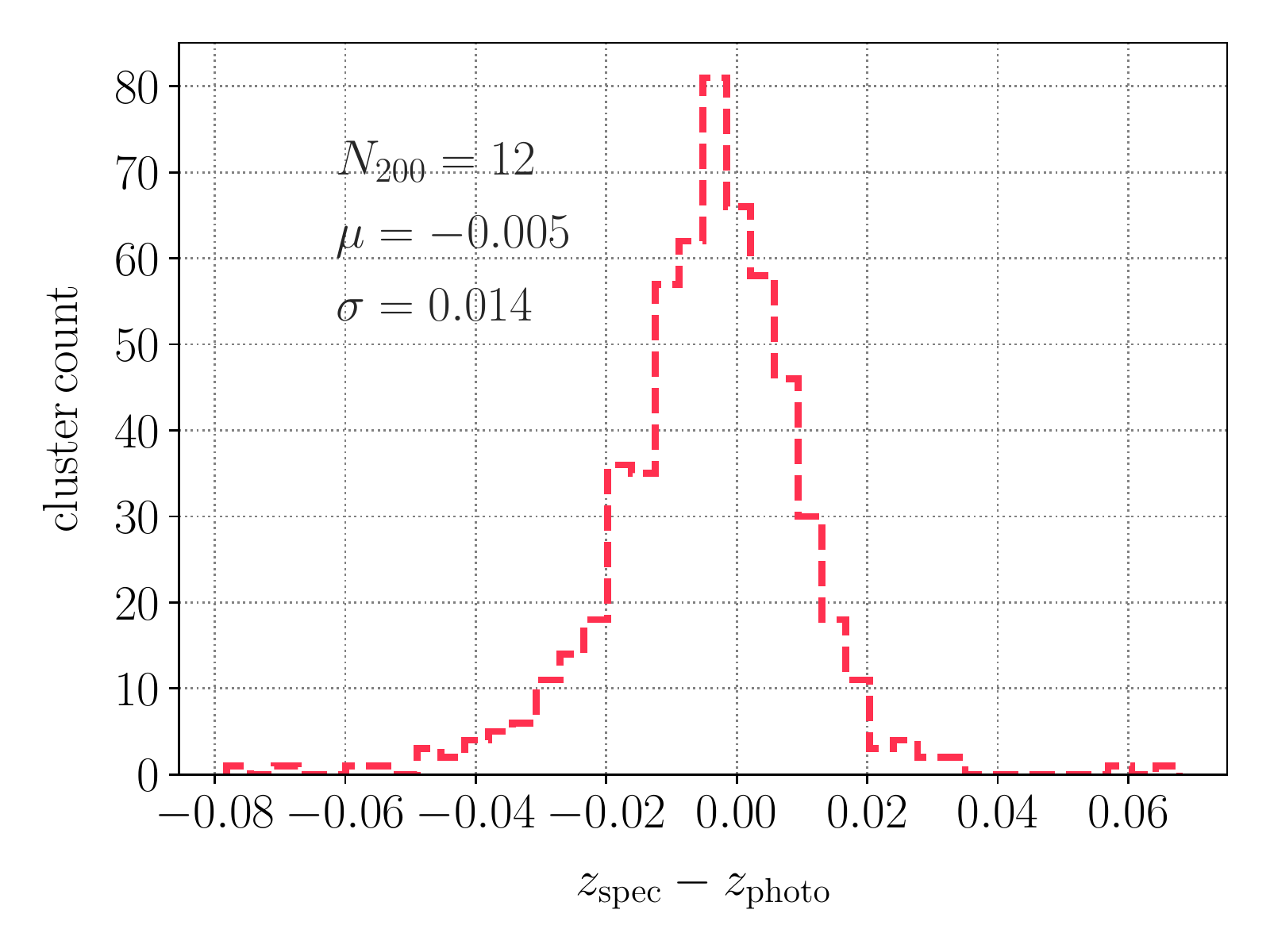}
   \caption{Histograms of differences between BCG spectroscopic redshift and cluster photometric redshift for 2129 maxBCG clusters below $M_{200}=0.85\times10^{14}\Msun$. We separate the clusters into three bins of richness $N_{200}=10$ (top, left panel), $N_{200}=11$ (top, right panel), $N_{200}=12$ (bottom panel) but detect no dependence of photo-z error on cluster richness. For clarity, the mean $\mu$ and standard deviation $\sigma$ of each histogram are additionally shown. 
      \label{fig:photoz_error}}
\end{figure}

To account for uncertainties in $v^{\LOS, s}_{L, i}$ induced by photometric redshift error (see left panel of \reffig{photoz_error}), we introduce an additional sampling step. Specifically, we generate a sample of $N_r$ realizations of maxBCG cluster positions in redshift space, in which we keep the redshifts of spec-z clusters fixed while sampling those of photo-z clusters as
\be
z^r_{\mathrm{photo}, i} = z^0_{\mathrm{photo}, i} + \delta z^r_i\,,
\ee
wherein $z^0_{\mathrm{photo}, i}$ is the \emph{fiducial} photometric redshift of cluster $i$ and
\be
\delta z^r_i \sim \N\left(0, \sigma_z\right).
\label{eq:photoz_gaussian_sampling}
\ee
Here, $\N\left(0, \sigma_z\right)$ is a zero-mean Gaussian with a standard deviation of $\sigma_z=0.015$. The latter value follows the scatter between spectroscopic and photometric redshifts of reference clusters $\sigma_z=\<\sigma[\zspec(N_{200})-\zphoto(N_{200})]\>$, averaged over three $N_{200}=[10,11,12]$ bins. As indicated in the right panel of \reffig{photoz_error}, this scatter shows no strong dependence on richness for the clusters considered in our analysis.
Note that we have assumed that there is no bias in the fiducial $\zphoto$, as the public version of maxBCG catalog is reportedly corrected for this effect \cite{Koester:2007}.
Additionally, we verified that our results are not significantly affected if we are to correct for a possible small bias of the order $\mu\simeq-0.005$ in $\zphoto$ (see \reffig{photoz_error}).

Note that when drawing from the Gaussian distribution in \refeq{photoz_gaussian_sampling}, we limit the range of $z^r_{\mathrm{photo}, i}$ in all redshift realizations to within $[0.05, 0.5]$. So our Gaussian distributions of photometric redshift error are actually truncated.

We then introduce an additional sum over all $N_r$ redshift realizations in \refeq{marginal_posterior},
\ba
& \P\left(\alpha|\{\D T_{\kSZ, i}/T_0\}\right) = 
\label{eq:double_sampling_posterior} \\
&\qquad \P\left(\alpha\right) \frac{1}{N_r} \sum_{r=1}^{N_r} \frac{1}{N_s} \sum_{s=1}^{N_s}  \prod_i \frac{1}{\sqrt{2\,\pi \, \sigma_i^2}} \exp\left\{-\frac{1}{2\sigma_i^2} \, \left (\frac{\D T_{\kSZ, i}}{T_0} +  \alpha \, \tau_i v^{\LOS, s r}_{L, i}/c \right)^2\right\}.
\nonumber
\ea
Note that, in the case of the AAP filter, for all redshift realizations, we keep the filter size in \refeq{double_sampling_posterior} fixed as $\theta_f=\varphi_f\,\theta^0_{200, i}$ where $\theta^0_{200, i}$ is computed from the fiducial values $z^0_{\mathrm{photo}, i}$.
Given the range in which the apparent size of selected maxBCG clusters varies (see right panel of \reffig{maxBCG_histograms}), this approximation does not affect our estimator in any significant way.

Since the two indices $s$ and $r$ in \refeq{double_sampling_posterior} are mathematically equivalent -- in the sense that they only appear in $v^{\LOS, s r}_{L, i}$ -- we can rewrite them as
\be
n \equiv \left\{s,\! r\right\}
\ee
and
\be
v^{\LOS,\ n}_{L, i} \equiv v^{\LOS, s}_{L, i}\left(z^r_{\mathrm{photo}, i}\right)
\ee
so that
\be
\P\left(\alpha|\{\D T_{\kSZ, i}/T_0\}\right) = \P\left(\alpha\right) \frac{1}{N} \sum_{n=1}^{N} \prod_i \frac{1}{\sqrt{2\,\pi \, \sigma_i^2}} \exp\left\{-\frac{1}{2\sigma_i^2} \, \left (\frac{\D T_{\kSZ, i}}{T_0} + \alpha \, x^n_i \right)^2\right\}\,,
\label{eq:sampling_posterior}
\ee
where $N=N_r\,N_s$ and $x^n_i\equiv\tau_i v^{\LOS, n}_{L, i}/c$.

The derivation of the estimator $\left\langle\alpha\right\rangle_n$ and its uncertainty $\sigma^2_\alpha$ is then almost identical to that in \refsec{kSZ:datamodel} and \refapp{kSZ:gaussianmixture}, only with $s$ replaced by $n$.


\section{Results}
\label{sec:kSZ:result}

Below we report the results of our measurement of the kSZ effect, imprinted by selected maxBCG clusters on the \smica2018 CMB map, in terms of the posterior mean $\<\alpha^{\varphi_f}\>_n$ -- over velocity and redshift realizations (cf. \refeq{alpha_ensemble_mean}) -- and the associated significance (cf. \refeq{significance}).
 
\subsection{Individual-scale signal measurements}
\label{sec:kSZ:individual_scale_AAP_result}
\begin{figure}
    \centerline{\resizebox{0.7\textwidth}{!}{\includegraphics*{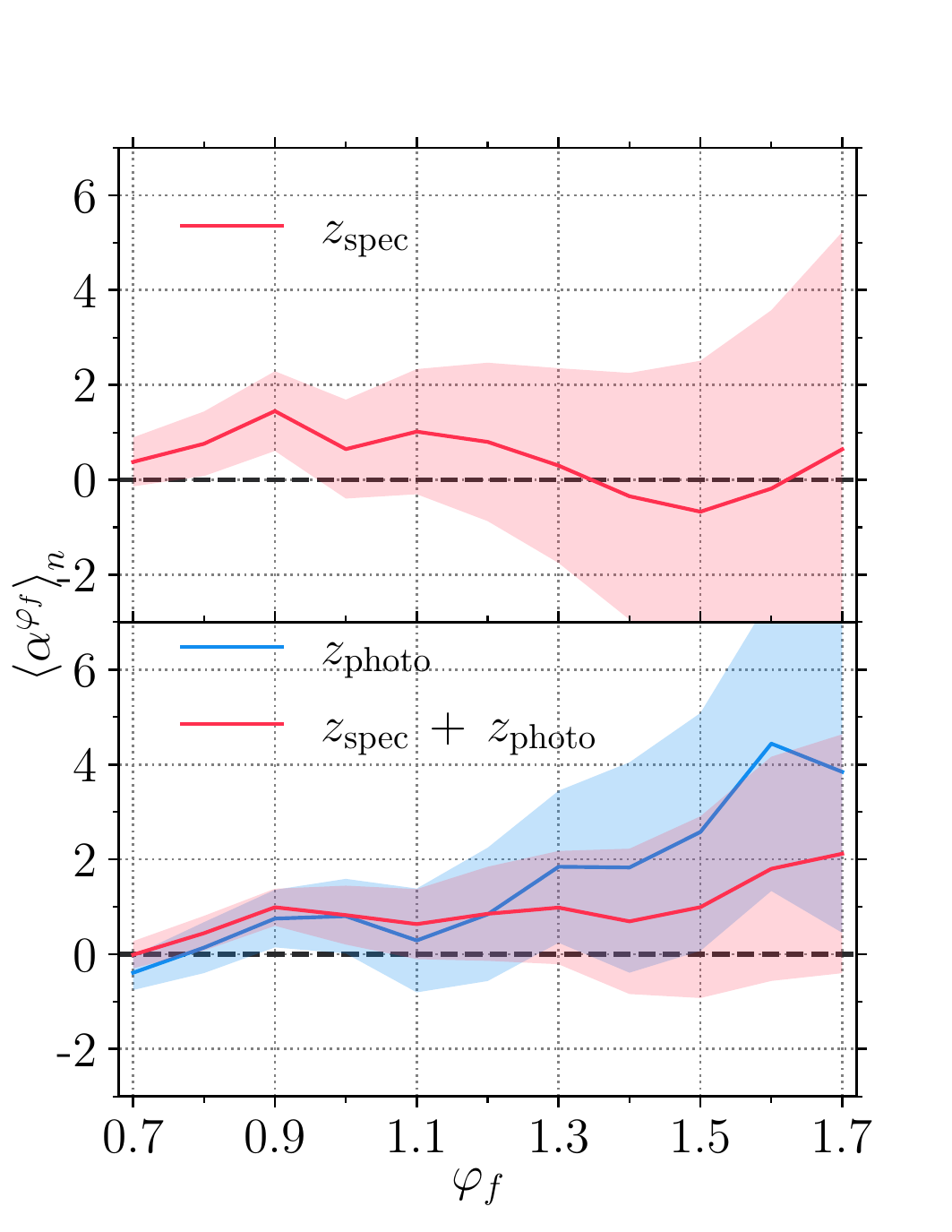}
    }}
    \caption{The posterior mean of the kSZ signal amplitude $\left\langle\alpha\right\rangle_n$ measured at different AAP filter scales $\varphi_f=[0.7,1.6]$ using spec-z (top panel, red) or photo-z (bottom panel, blue) and both sets (bottom panel, red). The shaded regions denote the corresponding $1\,\sigma$ uncertainties, which, in our case, including both uncertainties in CMB anisotropies and the reconstructed velocity field.}\label{fig:alpha_AAP_result}
  \end{figure}

We show in \reffig{alpha_AAP_result} the inferred value of $\<\alpha^{\varphi_f}\>_n$ as a function of the AAP filter scale $\varphi_f$, using spec-z set alone (top panel) or photo-z set and both sets (bottom panel).
We emphasize again that the $1\sigma$ uncertainty shown here as shaded band and reported in \reftab{AAP_SN} includes also uncertainties in the reconstructed large-scale velocity field.

As can be noted from \reffig{alpha_AAP_result}, we obtain consistent results between the two datasets spec-z and photo-z, which can be combined as shown in red in the bottom panel of \reffig{alpha_AAP_result}.
Note that the slightly larger uncertainty region of $\<\alpha\>_n$ measurement using spec-z set -- compared to that of $\<\alpha\>_n$ measurement using photo-z set -- is mostly caused by their limited quantity, as suggested by the ratio between the two uncertainties being approximately constant across all filter scales.
Both panels of \reffig{alpha_AAP_result} show that most of the information come from small scales. As the filter scale increases, the AAP estimate picks up more and more contribution from primary CMB anisotropies, as well as other large-scale sources of contamination, and quickly loose its constraining power.

In addition, we provide the significance at each filter scale for each dataset in \reftab{AAP_SN}. All sets show peaks of significance at $\varphi=0.9$, close to the angular cluster scale as one should expect. Further, for the photo-z set, the photo-z uncertainty shows a bigger impact at small filter sizes, which is also an expected behavior, since on large scales, the primary CMB is still the dominant noise source.

\begin{figure}[htbp!]
   \centering
   \includegraphics[width=0.495\textwidth]{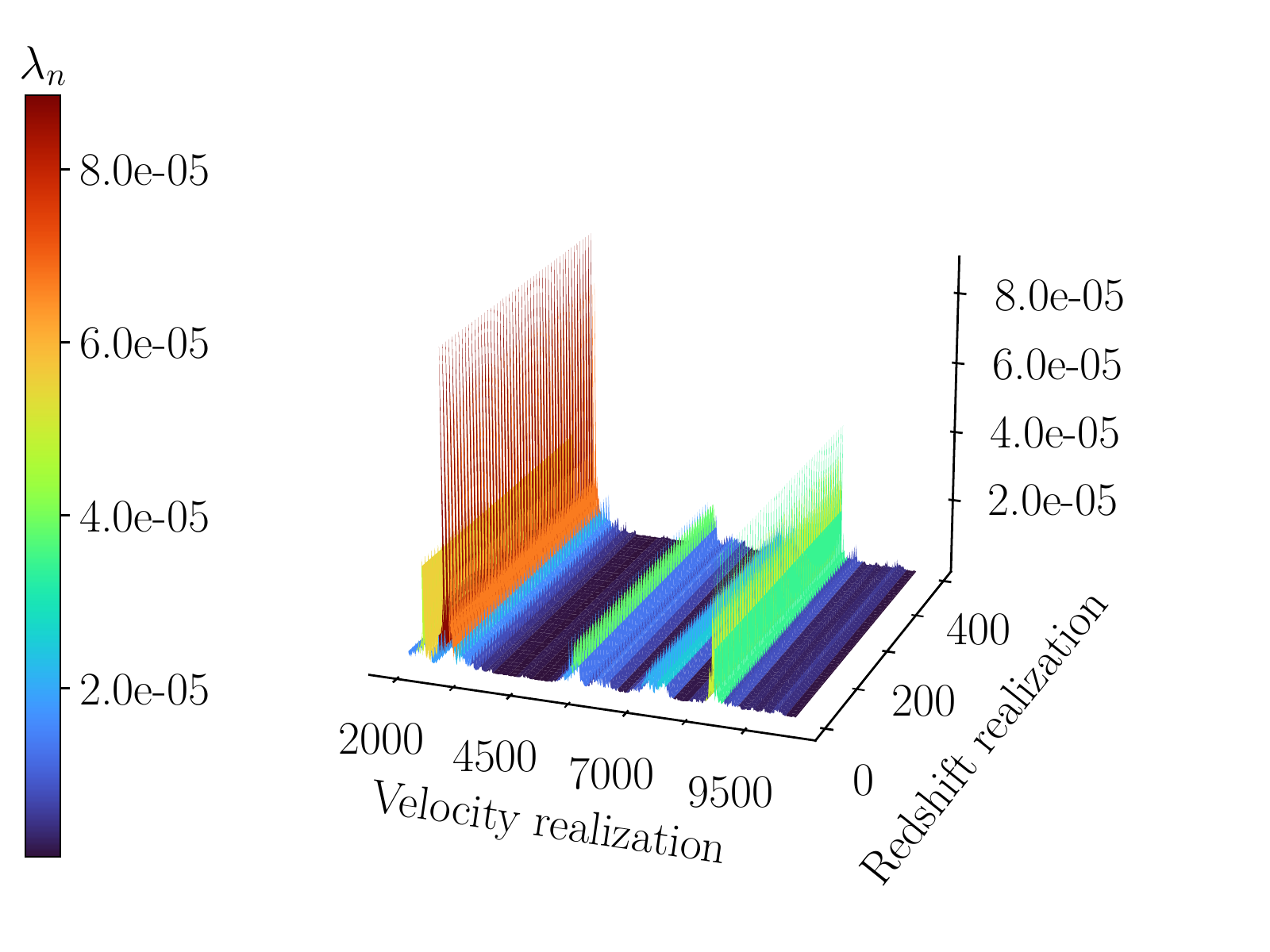}
   \includegraphics[width=0.495\textwidth]{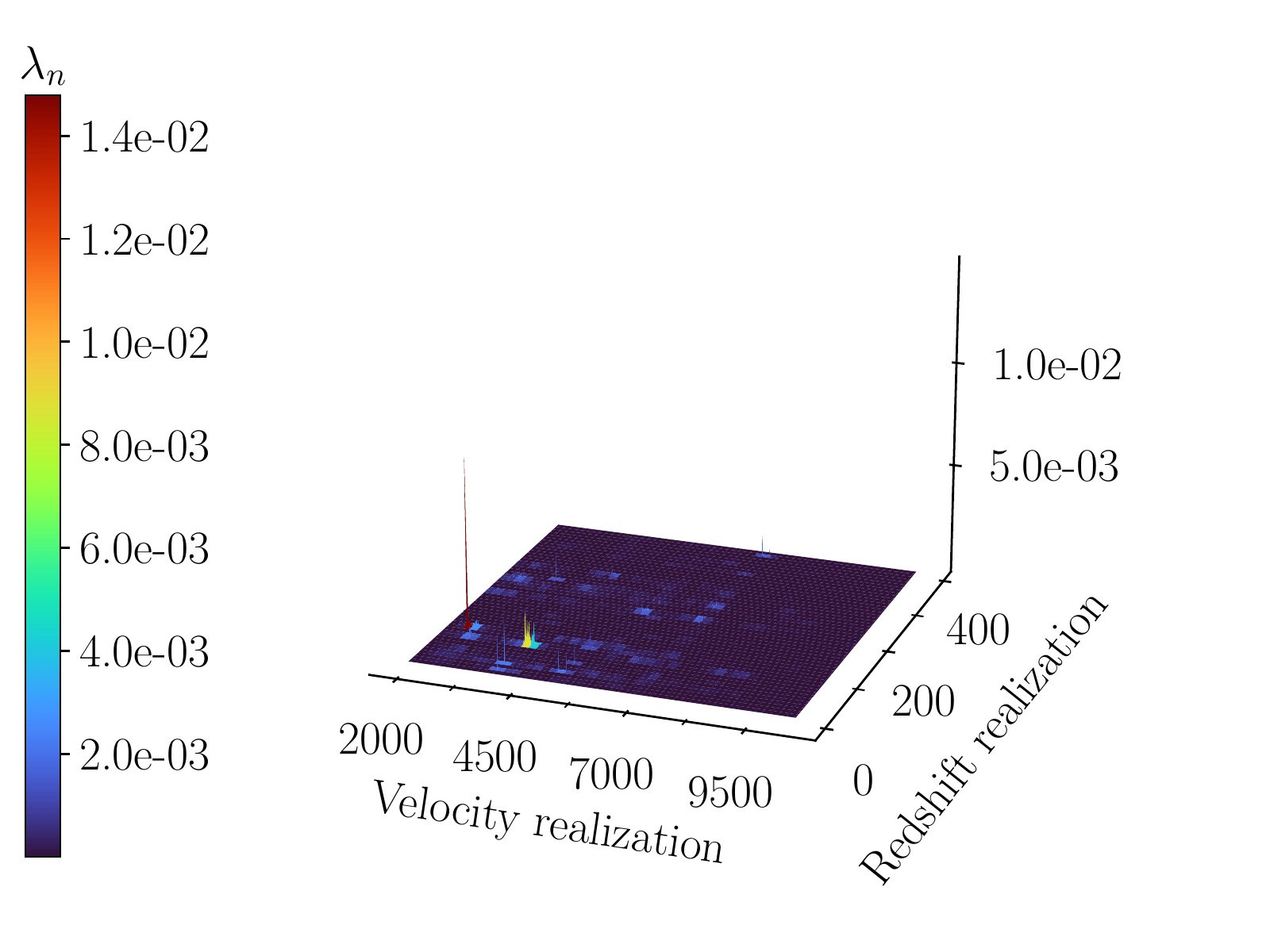}
   \includegraphics[width=0.495\textwidth]{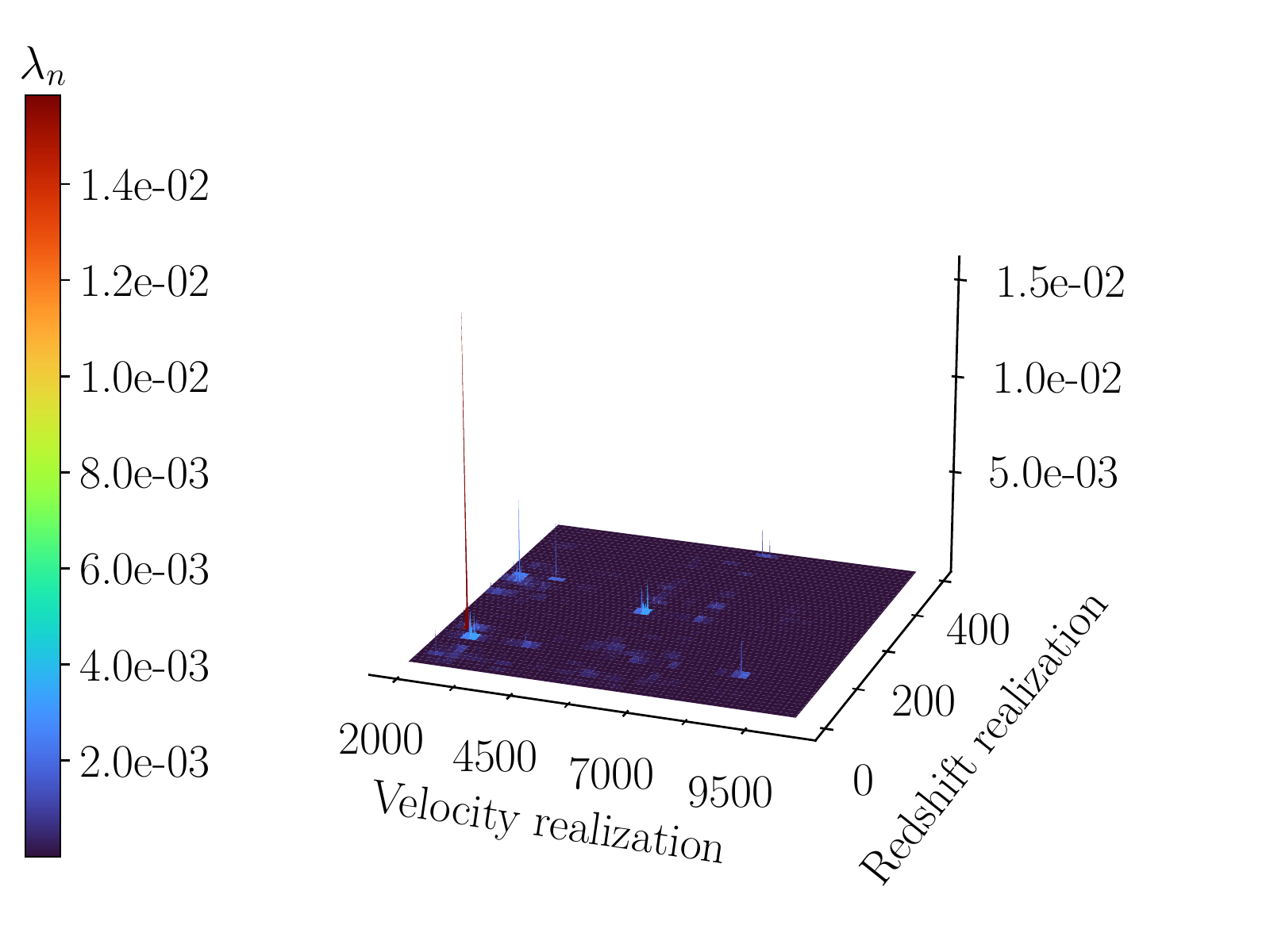}
   \caption{Distributions of the mixture weight $\lambda_n$ over 837 \borg-SDSS3 velocity realizations with mcmc identifier 2000-10360 (x-axes) and 400 redshift realizations (y-axes) for the cases of spec-z (left), photo-z (middle) and both datasets combined (right) at $\varphi=0.9$. Note that we do not sample the redshifts of clusters in the spec-z set, which is why the $\lambda_n$ are evenly distributed among redshift realizations (y-axes) in the leftmost column.\label{fig:lambda_AAP_1}}
\end{figure}

For illustration, we additionally show the mixture weights (cf. \refeq{lambda_s}) in \reffig{lambda_AAP_1}, since it can provide some insights on how the information in spec-z and photo-z sets are combined. One can see that the distributions of $\lambda_s$ in the three panels on the right of \reffig{lambda_AAP_1}, which show $\lambda_s$ for the combined set, are consistent with those of $\lambda_s$ in the leftmost and middle columns, which respectively show $\lambda_s$ for spec-z and photo-z set.
This implies that \refeq{sampling_posterior} consistently combines information from the spec-z and photo-z to simultaneously and correctly pick out the better-fitting \borg-SDSS3 samples and redshift realizations.

\addtolength{\tabcolsep}{8pt} 
\begin{table}[htbp!]
  \centering
  \renewcommand{\arraystretch}{1.2}
  \begin{tabular}{c c c c c c c}
  	\hline
	\hline
    $\varphi_f$ &\multicolumn{2}{c}{spec-z} &\multicolumn{2}{c}{photo-z} &\multicolumn{2}{c}{spec-z + photo-z}\\
    & $\<\alpha^{\varphi_f}\>_n$ & sign. & $\<\alpha^{\varphi_f}\>_n$ & sign. &  $\<\alpha^{\varphi_f}\>_n$ & sign. \\
    \hline  
    0.7 & $0.38\pm0.52$ & 0.76 & $-0.39\pm0.36$ & -1.06 & $-0.01\pm0.29$ & -0.09 \\
    0.8 & $0.76\pm0.68$ & 1.13 & $0.14\pm0.54$ & 0.27 & $0.44\pm0.37$ & 1.16 \\
    0.9 & $1.45\pm0.84$ & \textbf{1.71} & $0.75\pm0.61$ & \textbf{1.19} & $0.99\pm0.39$ & \textbf{2.21} \\
    1.0 & $0.65\pm1.04$ & 0.63 & $0.80\pm0.79$ & 0.98 & $0.82\pm0.62$ & 1.27 \\
    1.1 & $1.02\pm1.32$ & 0.78 & $0.29\pm1.09$ & 0.30 & $0.63\pm0.74$ & 0.85 \\
    1.2 & $0.80\pm1.67$ & 0.48 & $0.84\pm1.41$ & 0.61 & $0.85\pm0.99$ & 0.84 \\
    1.3 & $0.30\pm2.05$ & 0.15 & $1.84\pm1.60$ & 1.09 & $0.98\pm1.19$ & 0.81 \\
    1.4 & $-0.35\pm2.60$ & -0.14 & $1.83\pm2.22$ & 0.85 & $0.69\pm1.53$ & 0.46 \\
    1.5 & $-0.67\pm3.18$ & -0.21 & $2.58\pm2.51$ & 0.99 & $0.99\pm1.92$ & 0.51 \\
    1.6 & $-0.18\pm3.76$ & -0.05 & $4.44\pm3.11$ & 1.32 & $1.80\pm2.36$ & 0.76 \\
    1.7 & $0.65\pm4.58$ & 0.14 & $3.84\pm3.39$ & 1.08 & $2.12\pm2.52$ & 0.83 \\
    \hline
    \hline
  \end{tabular}
  \caption{
    The posterior mean of the kSZ signal amplitude $\<\alpha\>_n\pm1\sigma_\alpha$ measured at individual AAP filter scales, and the detection significance (cf. \refeq{significance}) are shown for, from left to right, spec-z, photo-z and both datasets. The maximum significance in each case is highlighted.
  }
   \label{tab:AAP_SN}
\end{table}
\addtolength{\tabcolsep}{-8pt}

\subsection{Combined signal measurement}
\label{sec:kSZ:multi_scale_AP_result}

\begin{figure}
    \centerline{\resizebox{0.7\textwidth}{!}{\includegraphics*{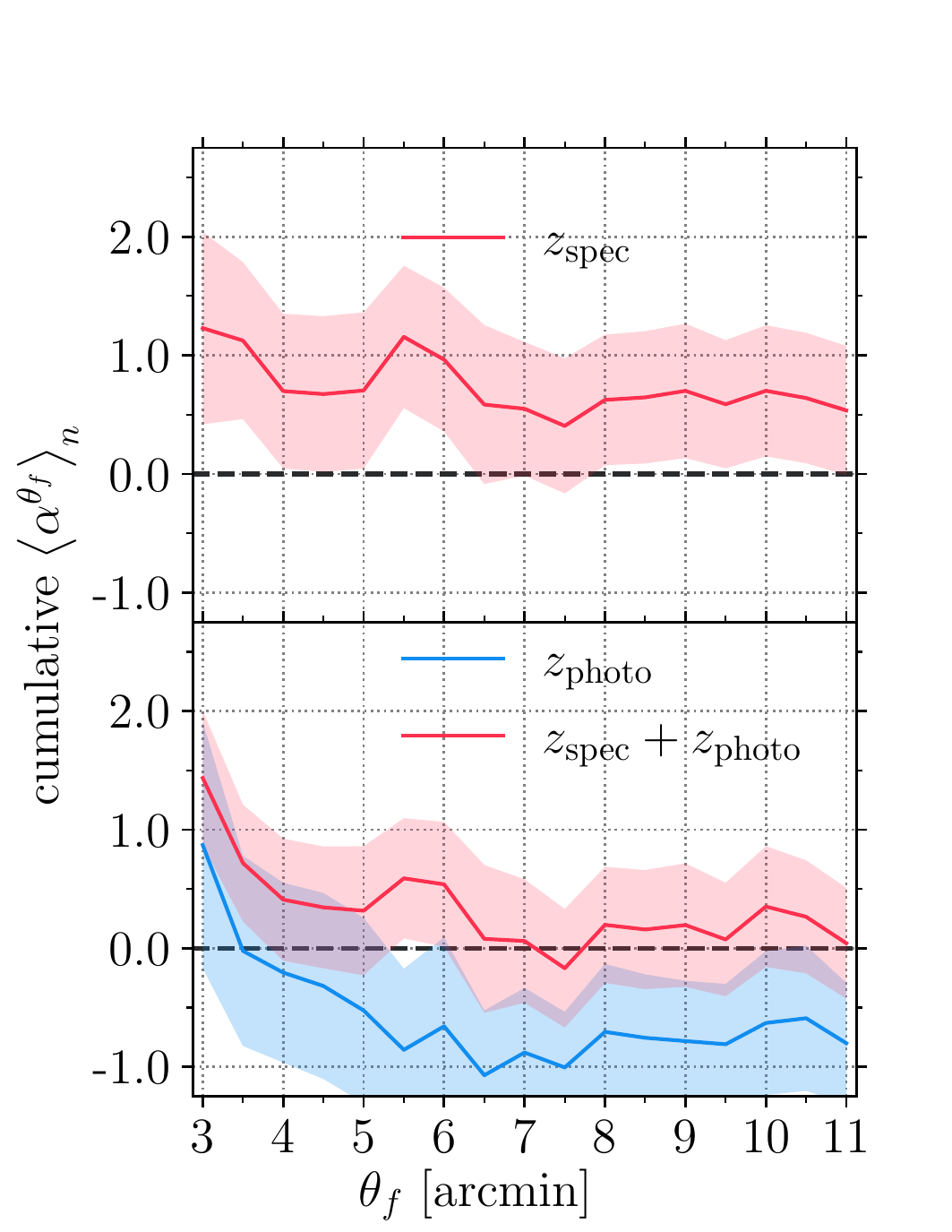}
    }}
    \caption{The posterior mean of the kSZ combined signal amplitude $\left\langle\alpha\right\rangle_n$ combining measurements at progressively larger AP filter sizes, $\theta_f=[3.0,11.0]$ arcmin, using spec-z (top panel, red) or photo-z (bottom panel, blue) and both sets (bottom panel, red). Going from left to right, each data point combines information from all previous points. The shaded regions denote the corresponding $1\,\sigma$ uncertainties, including both uncertainties in CMB anisotropies and the reconstructed velocity field.}\label{fig:alpha_AP_result}
  \end{figure}

We show in \reffig{alpha_AP_result} our measurements of combined signal amplitude $\<\alpha^{\theta_f}\>_n$ (cf. \refeq{alpha_ensemble_mean}), as a cumulative function of the AP filter radius $\theta_f$, for spec-z or photo-z set separately and both sets combined.

As can be seen from both cases of individual- and combined signal, the uncertainty in photometric redshift, when accounted for by double sampling, reduces the constraining power of the photo-z set, i.e. adding the clusters from the photo-z set does not substantially improve our significance. The most physically important factor is whether the photo-z error can be well-approximated by a \emph{symmetric} distribution, for example, the Gaussian, centered on the fiducial redshift. As clusters are virialized, collapsed objects, they are more likely to form in overdense regions. It is then reasonable to suspect that the photo-z fluctuations are not symmetrically distributed around the inferred value by the maxBCG algorithm, but rather biased towards overdense regions along the respective line of sight. Further detailed investigation is thus required to identify the optimal way to combine information from both spectroscopic and photometric data, subject to current constraints on computational resources. We defer such an investigation to future work.

The significance of the cumulative combined signal for each dataset is summarized in \reftab{AP_SN}. As in \reftab{AAP_SN}, most of the information is limited to filter sizes below and around the apparent size of maxBCG clusters in our samples. Above that scale, CMB noise severely limits our significance. This suggests that data from current CMB experiments such as ACTpol and SPT-3G with their much higher resolutions, $\sim1$ arcmin \cite{ACTPol:2016, SPTPol:2012}, can improve our significance significantly. On the other hand, even at the modest resolution of $\sim2-3$ arcmin, we expect a future experiment such as CMB-S4, with much lower instrumental noise, to also yield a significantly improved measurement. In both cases, the details of the gas profile would become important and one would almost certainly need to go beyond the Gaussian profile and the AP filter case considered here.
 
\addtolength{\tabcolsep}{8pt} 
\begin{table}[htbp!]
  \centering
  \renewcommand{\arraystretch}{1.2}
  \begin{tabular}{c c c c c c c}
  	\hline
	\hline
    $\theta_f$ [arcmin] &\multicolumn{2}{c}{spec-z} &\multicolumn{2}{c}{photo-z} &\multicolumn{2}{c}{spec-z + photo-z}\\
    & $\<\alpha^{\varphi_f}\>_n$ & sign. & $\<\alpha^{\varphi_f}\>_n$ & sign. &  $\<\alpha^{\varphi_f}\>_n$ & sign. \\
    \hline
    3.0 & $1.23\pm0.81$ & 1.51 & $0.87\pm1.04$ & \textbf{0.81} & $1.44\pm0.57$ & \textbf{2.09} \\
    3.5 & $1.12\pm0.66$ & 1.67 & $-0.02\pm0.80$ & -0.01 & $0.72\pm0.49$ & 1.39 \\
    4.0 & $0.70\pm0.65$ & 1.07 & $-0.21\pm0.76$ & -0.26 & $0.41\pm0.51$ & 0.79 \\
    4.5 & $0.67\pm0.66$ & 1.03 & $-0.32\pm0.78$ & -0.40 & $0.34\pm0.51$ & 0.67 \\
    5.0 & $0.70\pm0.66$ & 1.07 & $-0.53\pm0.78$ & -0.66 & $0.32\pm0.54$ & 0.60 \\
    5.5 & $1.16\pm0.60$ & \textbf{1.88} & $-0.86\pm0.68$ & -1.15 & $0.59\pm0.51$ & 1.12 \\
    6.0 & $0.96\pm0.61$ & 1.56 & $-0.66\pm0.75$ & -0.90 & $0.54\pm0.53$ & 0.98 \\
    6.5 & $0.58\pm0.67$ & 0.91 & $-1.07\pm0.55$ & -1.63 & $0.08\pm0.62$ & 0.11 \\
    7.0 & $0.55\pm0.56$ & 0.98 & $-0.88\pm0.55$ & -1.41 & $0.06\pm0.52$ & 0.12 \\
    \hline
    \hline
  \end{tabular}
   \caption{The posterior mean of the kSZ combined signal amplitude $\<\alpha\>_n\pm1\sigma_\alpha$ measured at increasing maximum AP filter size, and the corresponding detection significance (cf. \refeq{significance}) are shown for spec-z, photo-z and both datasets (left to right). The maximum significance in each case is highlighted. Note that, for readability we only list here results for $\theta_f=[3,7]$ arcmin.}
   \label{tab:AP_SN}
\end{table}
\addtolength{\tabcolsep}{-8pt}

\section{Null tests for systematics}
\label{sec:kSZ:test}

In this section, we further assert the significance of our measurement by performing two null tests on mock data in which we
\begin{enumerate}
	\item shuffle the position of the clusters in our analysis, and
	\item replace the \smica2018 by a set of 300 \smica-like mock maps, also taken from Planck 2018 data release \cite{Planck:2018IV}.
\end{enumerate}
Since the photo-z set adds very little information, in the tests below, we will only focus on the spec-z set.
We have found that the significance computed from \refeq{significance} can be rather well approximated by the simple ratio S/N=$\<\alpha^{\varphi_f}\>_n/\sigma_{\alpha}$. Below, we employ this approximation to ease the numerical computation of the significance in these null tests.

\subsection{Null tests: Individual-scale signal}
\label{sec:kSZ:individual_scale_AAP_test}

For the case of individual-scale measurements using the AAP filter, we measure a signal amplitude consistent with zero for the spec-z set with cluster positions being shuffled, as can be seen in the left panel of \reffig{shuffle_AAP_and_AP_null_test}.
When applying our pipeline on the set of 300 \smica2018 simulations (including CMB and instrumental noise) provided by Planck \cite{Planck:2018IV}, we also recover S/N ratios consistent with zero-mean Gaussian distributions. We show in the top panels of \reffig{histogram_mock_AAP_and_AP_null_test} the histograms of S/N for increasing individual filter scale from $\varphi=0.9$ to $\varphi=1.1$.
The histogram of $\varphi=0.9$ (top row, left panel) appears consistent with our reported significances of $\simeq1.7$ for the spec-z dataset (cf. \reftab{AAP_SN}).


\subsection{Null tests: combined signal}
\label{sec:kSZ:multi_scale_AP_test}

\begin{figure*}
    \centerline{\resizebox{\hsize}{!}{\includegraphics*{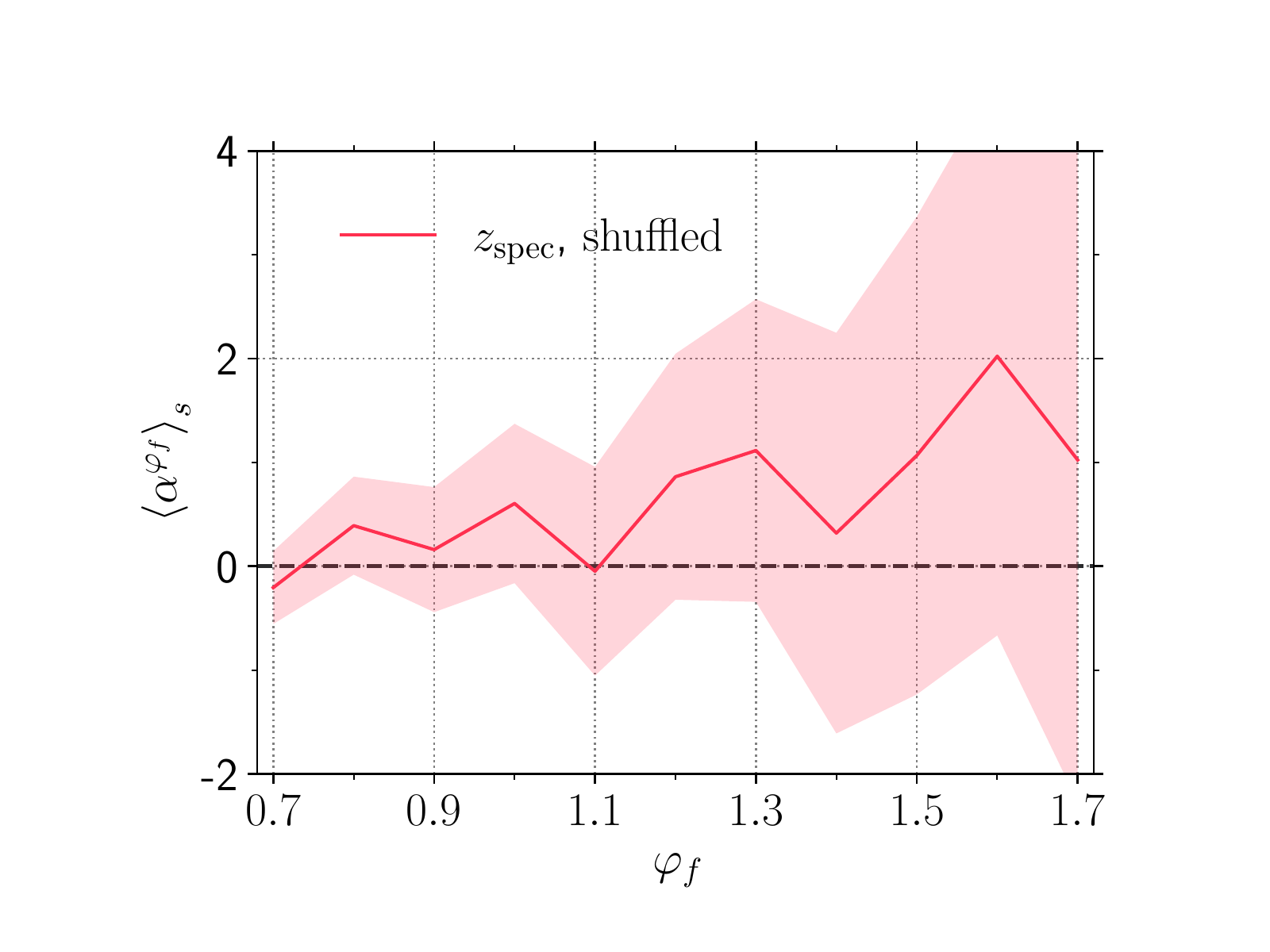} \,
        \includegraphics*{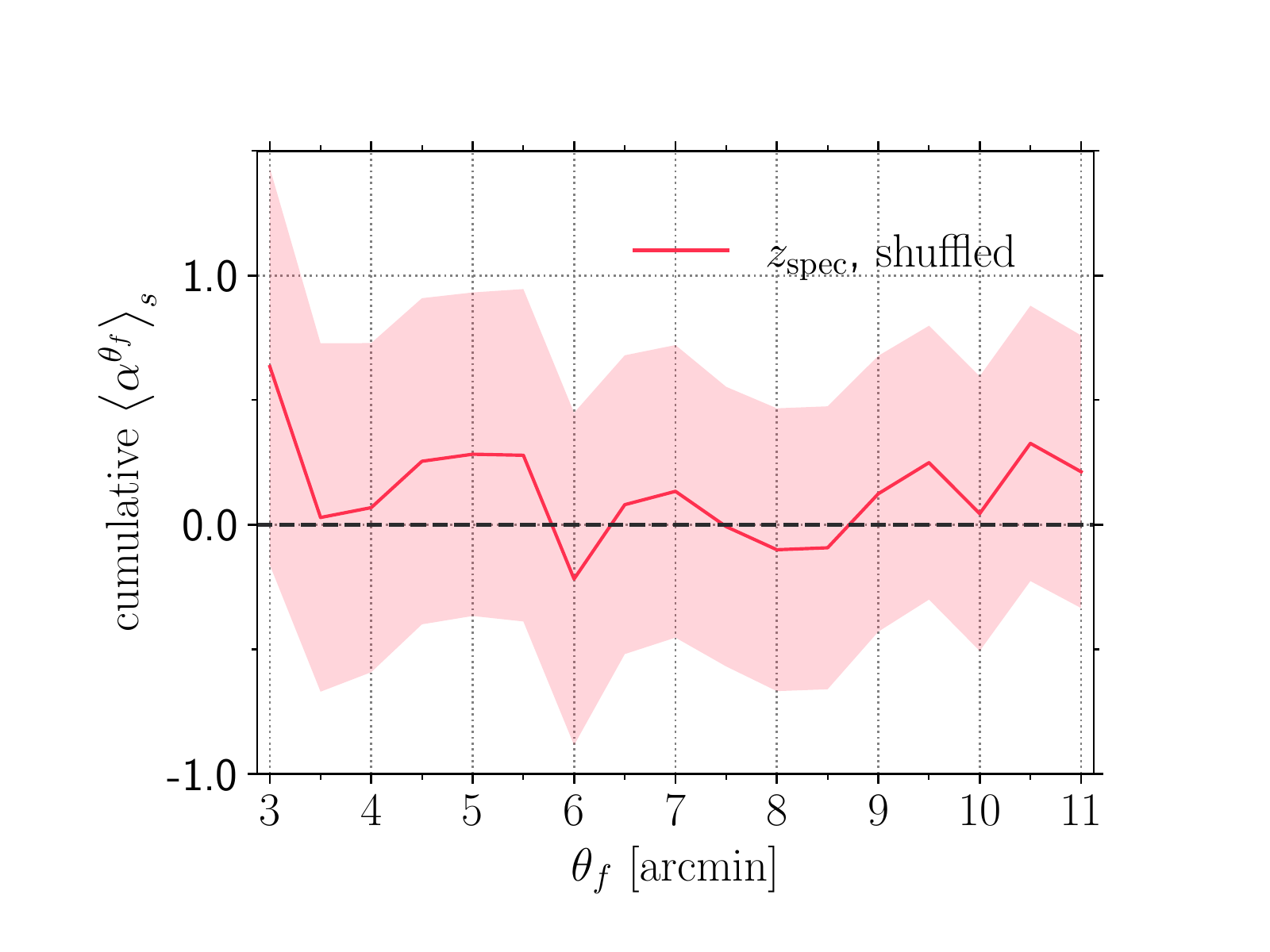}
        }}
    \caption{\textit{Left panel:} Individual-scale signal amplitude measured using spec-z sample but with sky positions of the clusters shuffled, plotted as a function of AAP filter scale.
       \textit{Right panel:} combined signal amplitude measured using spec-z set but with sky positions of the clusters shuffled, plotted as a cumulative function of AP filter sizes.}\label{fig:shuffle_AAP_and_AP_null_test}
\end{figure*}

\begin{figure*}%
\begin{subfigure}{0.31\textwidth}
\includegraphics*[width=\linewidth]{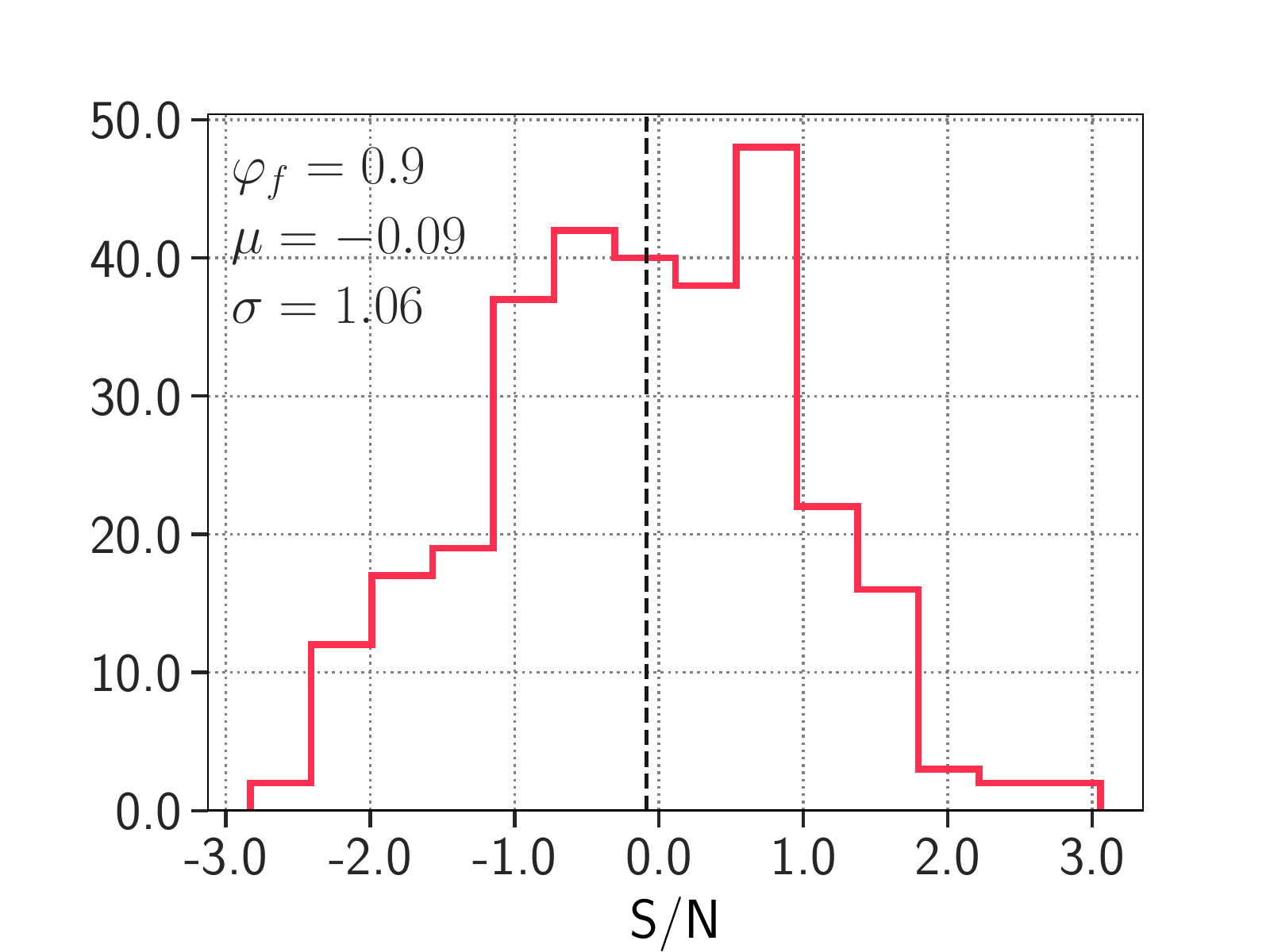}
\end{subfigure}
\begin{subfigure}{0.31\textwidth}
\includegraphics*[width=\linewidth]{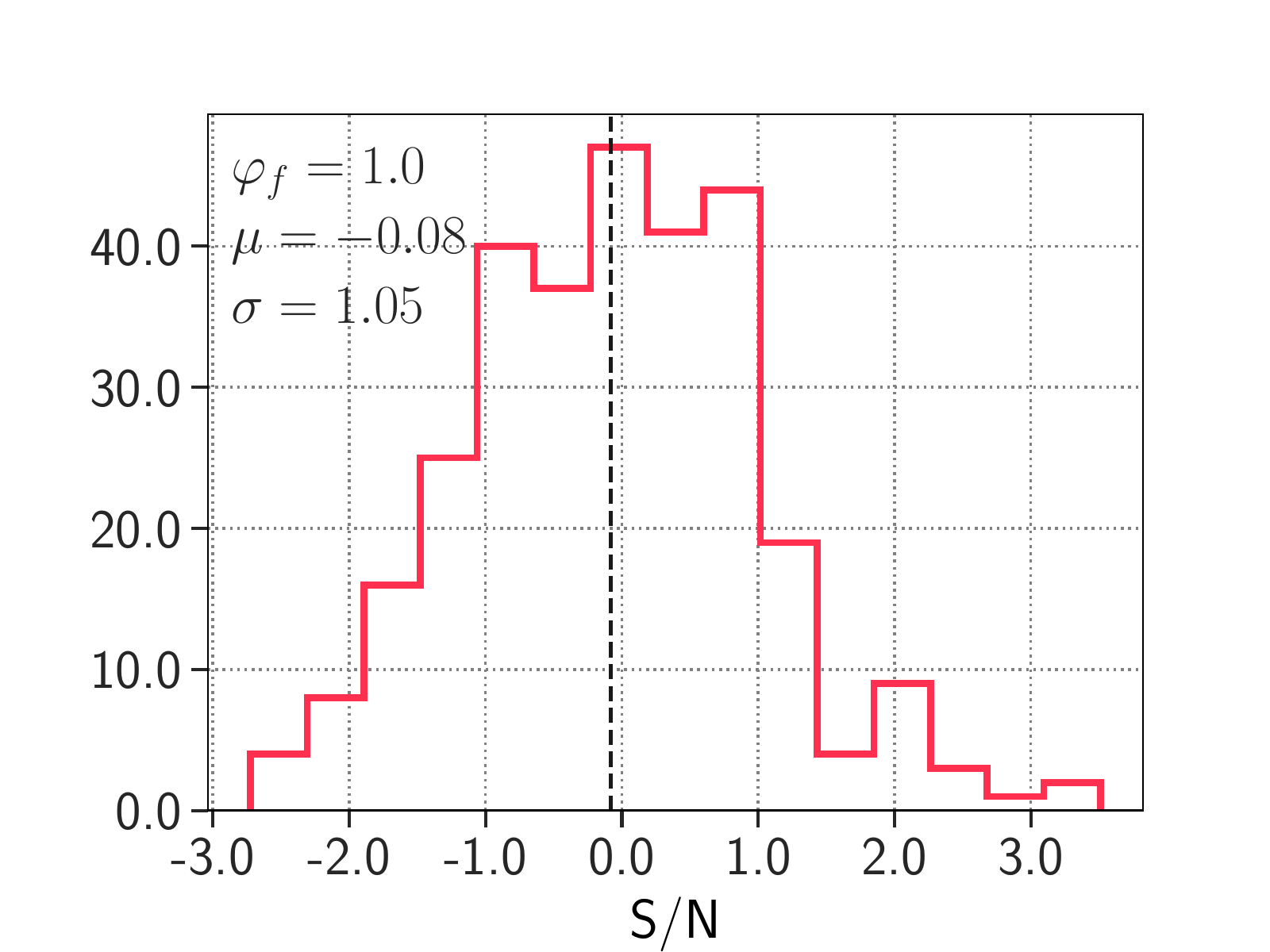}
\end{subfigure}
\begin{subfigure}{0.31\textwidth}
\includegraphics*[width=\linewidth]{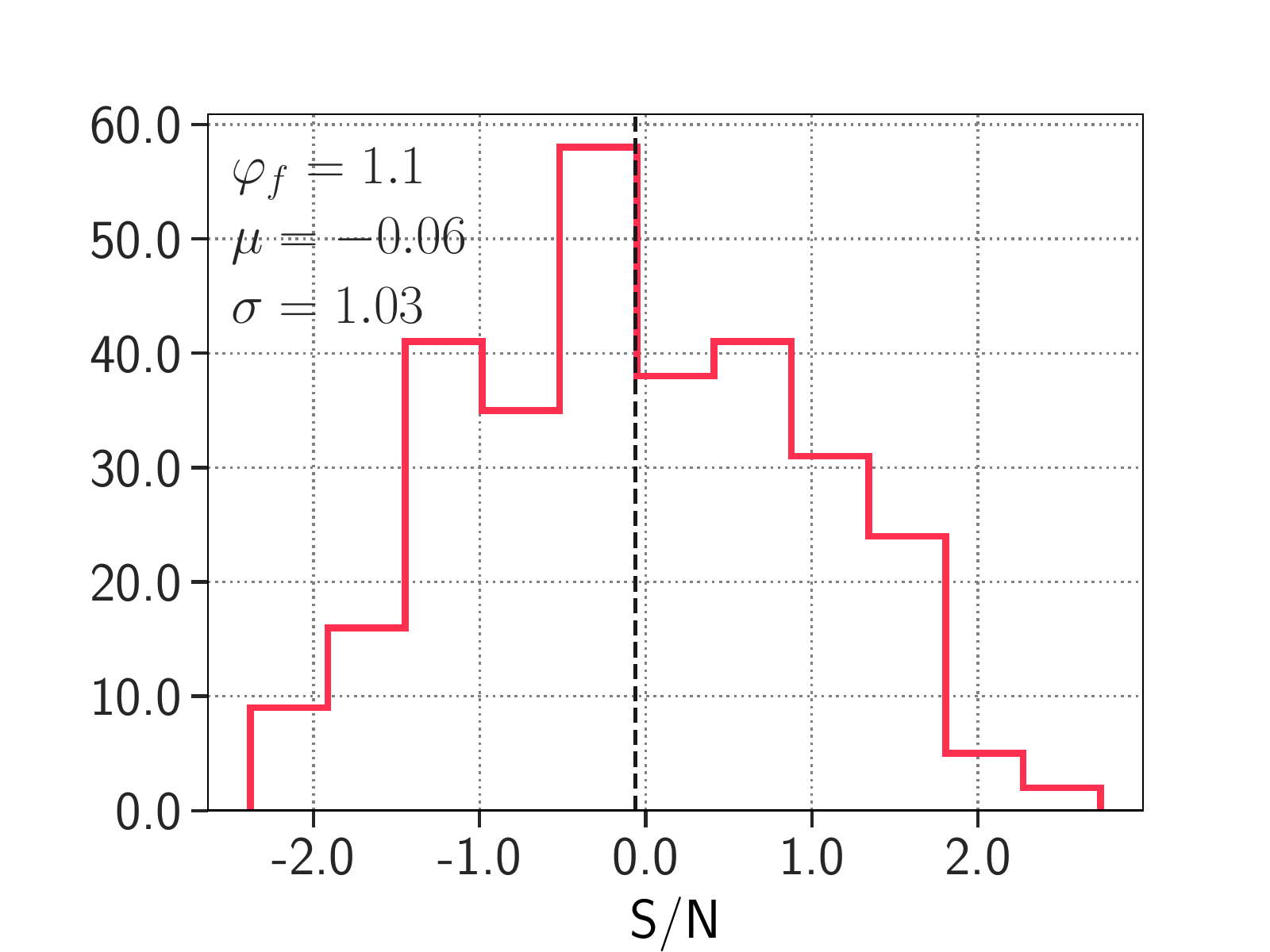}
\end{subfigure}

\begin{subfigure}{0.31\textwidth}
\includegraphics*[width=\linewidth]{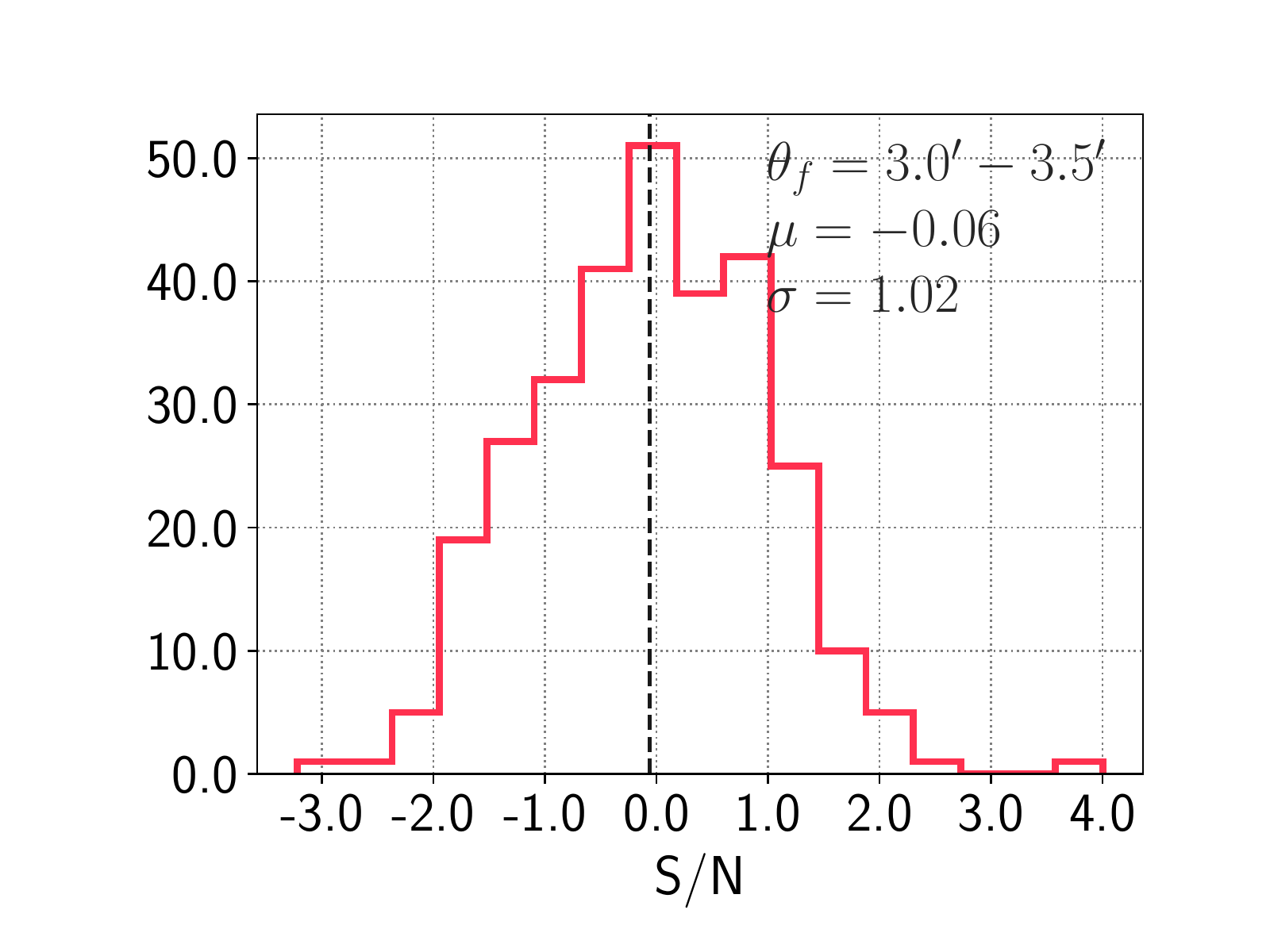}
\end{subfigure}
\begin{subfigure}{0.31\textwidth}
\includegraphics*[width=\linewidth]{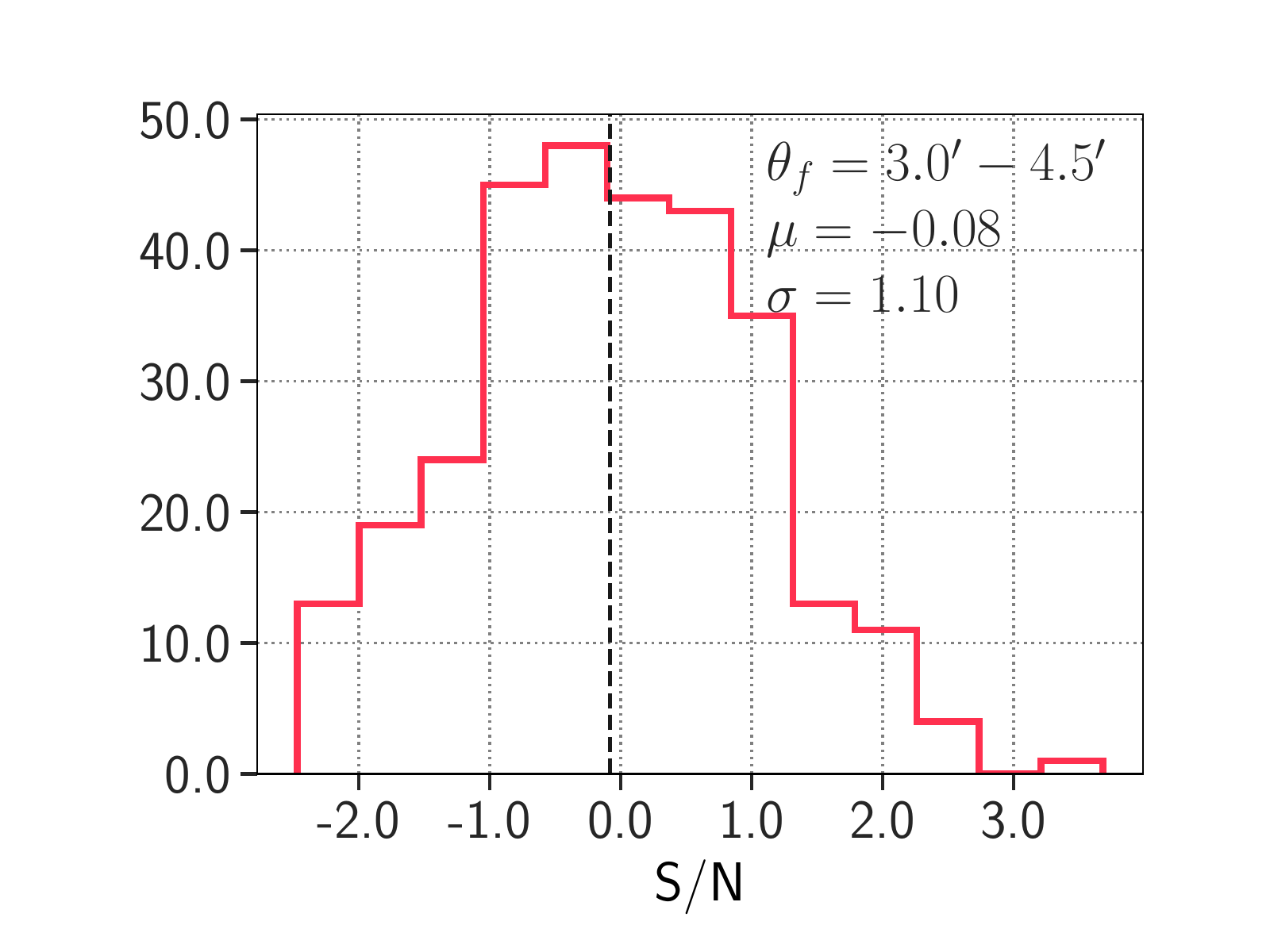}
\end{subfigure}
\begin{subfigure}{0.31\textwidth}
\includegraphics*[width=\linewidth]{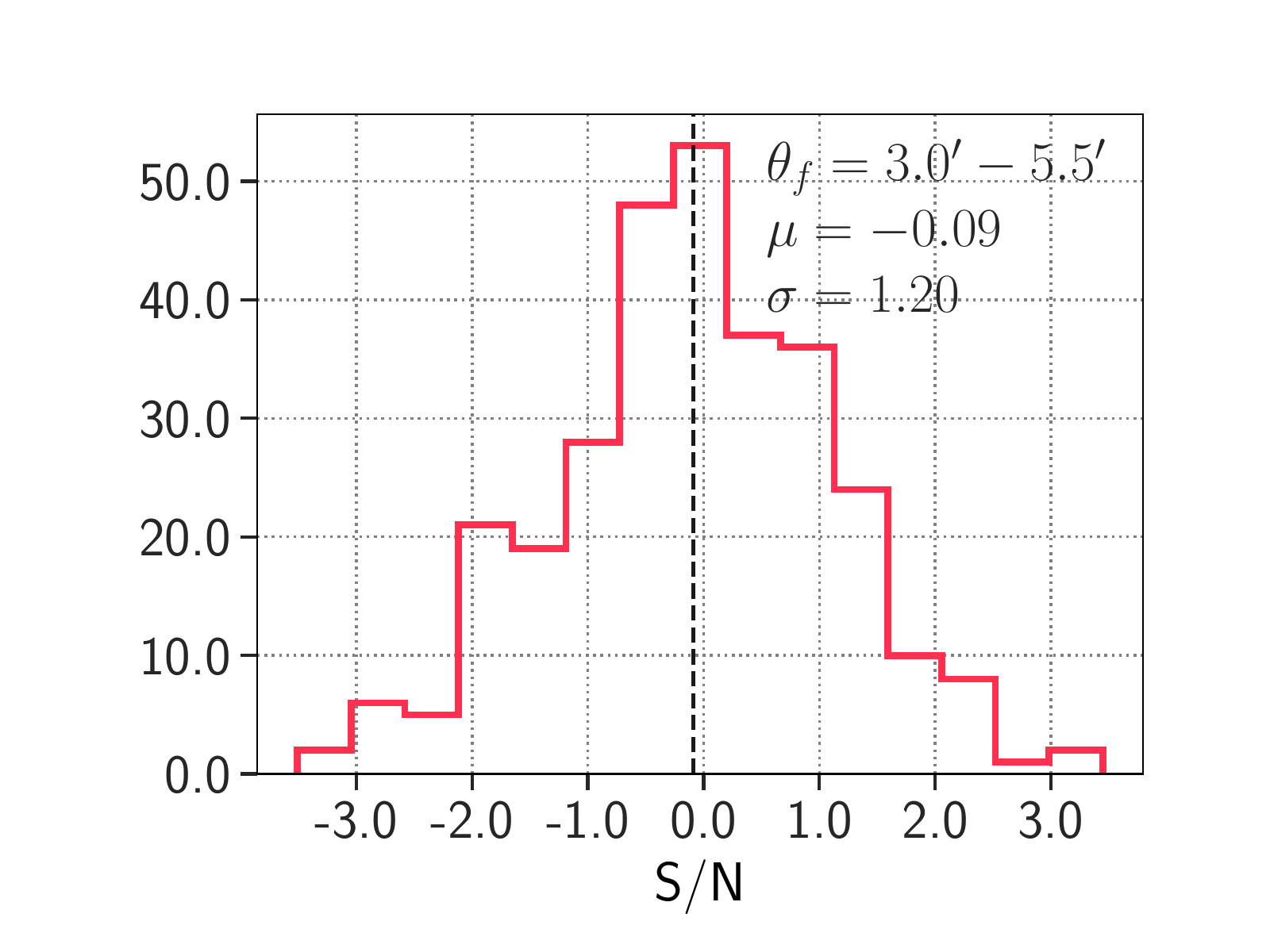}
\end{subfigure}

\caption{\textit{Top:} Histograms of the S/N of individual-scale signal for spec-z set, measured at $\varphi_f=[0.9,1.0,1.1]$, from 300 simulations of \smica2018 (including CMB and \smica-like instrumental noise). The vertical dashed line presents the mean in each histogram. For clarity, the mean $\mu$ and standard deviation $\sigma$ of each histogram are also shown in each plot.
\textit{Bottom:} Histograms of the S/N of multi-scale signal for spec-z set, measured cumulatively between $\theta_f=3.0$ and $\theta_f=5.5$ arcmin, from 300 simulations of \smica2018 (including CMB and \smica-like instrumental noise). The vertical dashed line presents the mean in each histogram. For clarity, the mean $\mu$ and standard deviation $\sigma$ of each histogram are also reported. The y-axis of each histogram displays the \borg-SDSS3 velocity realization count.} \label{fig:histogram_mock_AAP_and_AP_null_test}
\end{figure*}

For the case of cumulative multi-scale measurements using the AP filter, we also measure a cumulative signal amplitude consistent with zero for spec-z set with cluster positions being shuffled, as shown in \reffig{shuffle_AAP_and_AP_null_test}.
Our null test using the Planck simulations also recovers S/N ratios consistent with zero-mean Gaussian distributions. We show in \reffig{histogram_mock_AAP_and_AP_null_test} the histograms of S/N for cumulative multi-scale from $\theta_f=3.0-3.5$ arcmin to $\theta_f=3.0-5.5$ arcmin.
The histogram of $\theta_f=3.0-5.5$ arcmin (bottom row, right panel) again shows that our reported significance of $\simeq1.9$ for the spec-z set (cf. \reftab{AP_SN}) is consistent with the 95\% confidence interval.

\section{Discussion and conclusion}
\label{sec:kSZ:conclusion}

kSZ measurements have shown promising potential to become a new probe for both ionized baryons and Dark Energy with the next generation of CMB experiments. It is thus important to reduce systematic uncertainties in cosmology inference from future kSZ measurements. So far, one of the often neglected sources of systematics is that in the reconstructed velocity.  
Using a systematic-free ensemble of inferred velocity fields within the SDSS3-BOSS volume, in \refsec{kSZ:datamodel}, we have developed a robust likelihood for the kSZ effect that, for the first time, accounts for uncertainties in the velocity reconstruction process into the final uncertainty on the measured signal. As such, the significance of our kSZ measurement can be thought of as a rigorously conservative estimate.
It is worth pointing out that extending or optimizing our framework for specific studies is relatively straightforward. For example, one can assume specific cluster gas profiles and adopt more sophisticated filtering techniques. Another advantage of our approach is that a prior on $\alpha$ can be easily introduced for cases wherein the cluster gas profile is known from complementary measurements, e.g. X-ray, tSZ, etc.

In addition, we would like to remark on one of the main concerns regarding our approach: the numerical expense of including the uncertainty from velocity reconstruction through forward modeling. Here, the major bottleneck is, however, the initial conditions reconstruction process in \cite{Lavaux:2019} of which this work is built on. A particular question is, to which resolution can this method achieve with larger volume surveys? To this end, it is worth pointing out that there are several performance improvements have been developing and implementing within the \borg{} framework, including the flexibility of having different resolutions for the initial and evolved matter density field. Technically, one can then achieve the same resolution with, for example, a survey of two-to-four times effective volume size, while having to sample only the same parameter space (of the initial conditions and bias parameters). All things considered, the advantage of including the uncertainty in velocity reconstruction (so as to not bias any inferred quantity, for example, the optical depth, measured from the kSZ effect) far outweighs the disadvantage of numerical complexity it introduces. This is especially important as recent and upcoming kSZ measurements keep improving their S/N ratios \cite{Schaan:2020, Amodeo:2020}.

We apply our method to measure the kSZ signal, imprinted by large-scale bulk flow of selected maxBCG clusters, in the \smica2018 CMB map. We observe, as reported in \refsec{kSZ:result}, moderate evidence of the signal at $\simeq2\sigma$ -- including velocity uncertainty. In this simple demonstration, the sensitivity of our kSZ measurement is limited by several factors, including the simple AP filter approach, the typical cluster scale being close to the resolution of the \smica2018 map, and the current level of instrumental noise. Yet, we find that ignoring systematic uncertainty in the reconstructed velocity could already lead to a significant bias in the kSZ measurement (cf. \refeq{alpha_ensemble_variance} and \reffig{sigma_single_vs_ensemble_VLOS}). It would be interesting to study how this would affect kSZ measurements using more sophisticated filtering methods, for example, the matched-filter approach, and data from higher resolution and sensitivity CMB experiments, e.g. CMB-S4 \cite{CMB-S4:2016}, SPT-3G \cite{SPT3G:2014}.

For the sole purpose of kSZ detection, another important factor is the sheer number of clusters. In fact, our main dataset, the maxBCG spec-z sample, is quite limited in number at only 908 clusters. To overcome this issue, we have explored in \refsec{kSZ:photoz_model} the possibility of including clusters with only photometric redshifts, the maxBCG photo-z sample, by sampling each clusters' redshift from a symmetric Gaussian distribution centered on the fiducial value. However, we only saw a modest increase in detection significance when including photometric clusters. We leave a detailed investigation on whether the true underlying distribution of redshift fluctuations is Gaussian, and whether more information can be extracted from photometric clusters, for future work.

In this work, we have ignored the thermal SZ (tSZ) signal from clusters, and opted to remove the most massive clusters from the analysis. 
One could also think of simultaneously modeling and measuring both kSZ and tSZ signals for each cluster, so that there is no need for the removal of massive clusters. Nothing, in principle, prevents this simple extension of the data model in \refeq{datamodel}. In fact, we plan to pursue this approach in a follow-up study.

\section*{Data and code availability}
 For the BOSS-SDSS3 reconstruction data, please contact \url{https://www.aquila-consortium.org/contact/}. The numerical implementation of the kSZ likelihood presented in this paper is available upon request to the corresponding author. Public releases of both will be available in the future, accompanying follow-up works.

   \FloatBarrier

   
   \acknowledgments

   We would like to thank Eiichiro Komatsu, Jochen Weller, Franz Elsner, Alex Barreira, Giovani Cabass, Seunghwan Lim, Emmanuel Schaan, Naonori Sugiyama, Daisuke Nagai, Shun Saito, as well as the members of the Aquila Consortium for inspiring and helpful discussions.
   MN and FS acknowledge support from the Starting Grant (ERC-2015-STG 678652) \enquote{GrInflaGal} of the European Research Council.
   GL acknowledges financial support from the ILP LABEX (under reference ANR-10-LABX-63) which is financed by French state funds managed by the ANR within the Investissements d'Avenir programme under reference ANR-11-IDEX-0004-02. This work was supported by the ANR BIG4 project, grant ANR-16-CE23-0002 of the French Agence Nationale de la Recherche.
   This work is done within the Aquila Consortium.\footnote{\url{https://aquila-consortium.org}}


   \appendix

   \section{tSZ contamination and cluster mass cut}
   \label{app:kSZ:masscut}

\begin{figure*}
    \centerline{\resizebox{\hsize}{!}{\includegraphics*{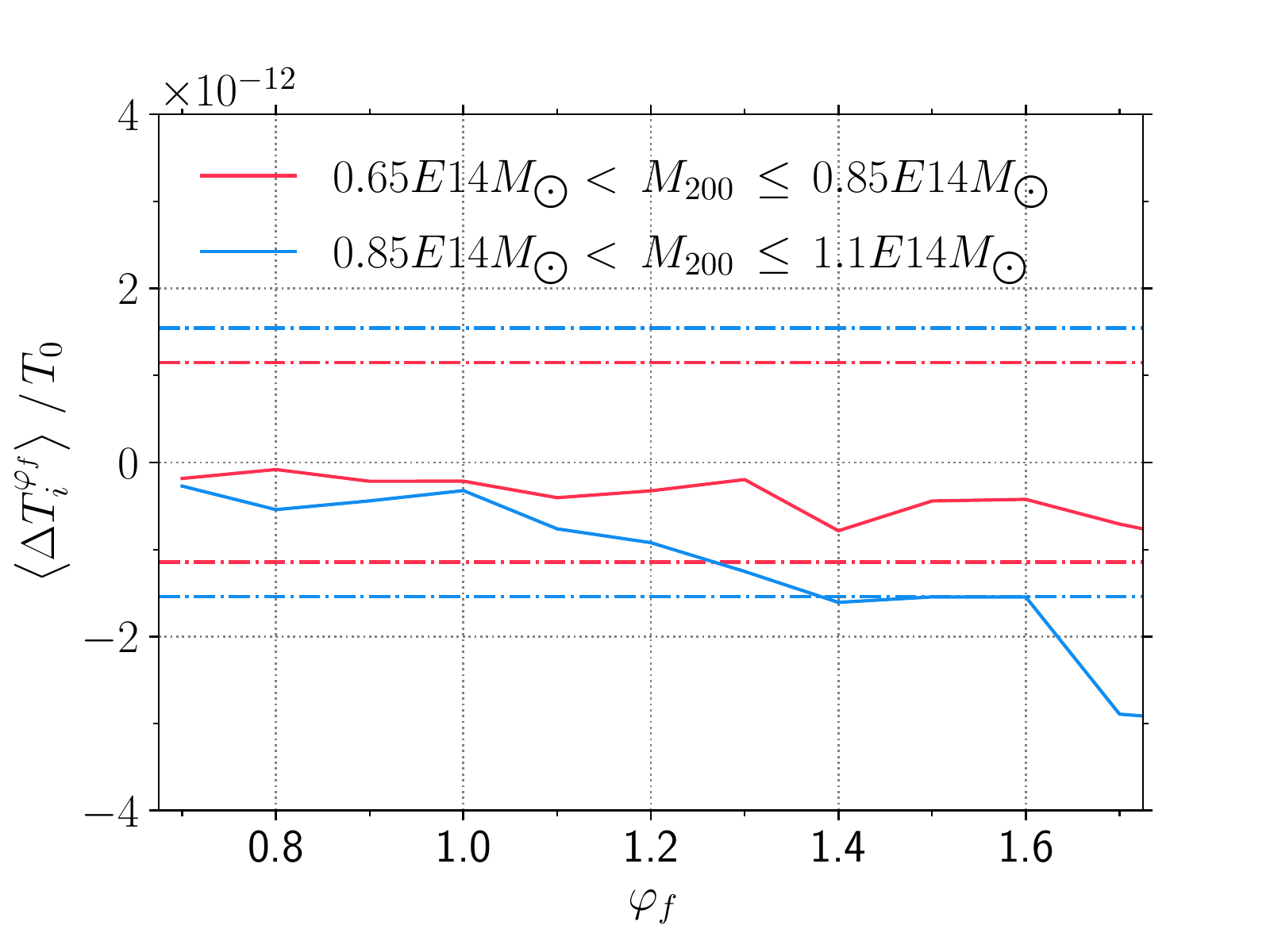} \,
        \includegraphics*{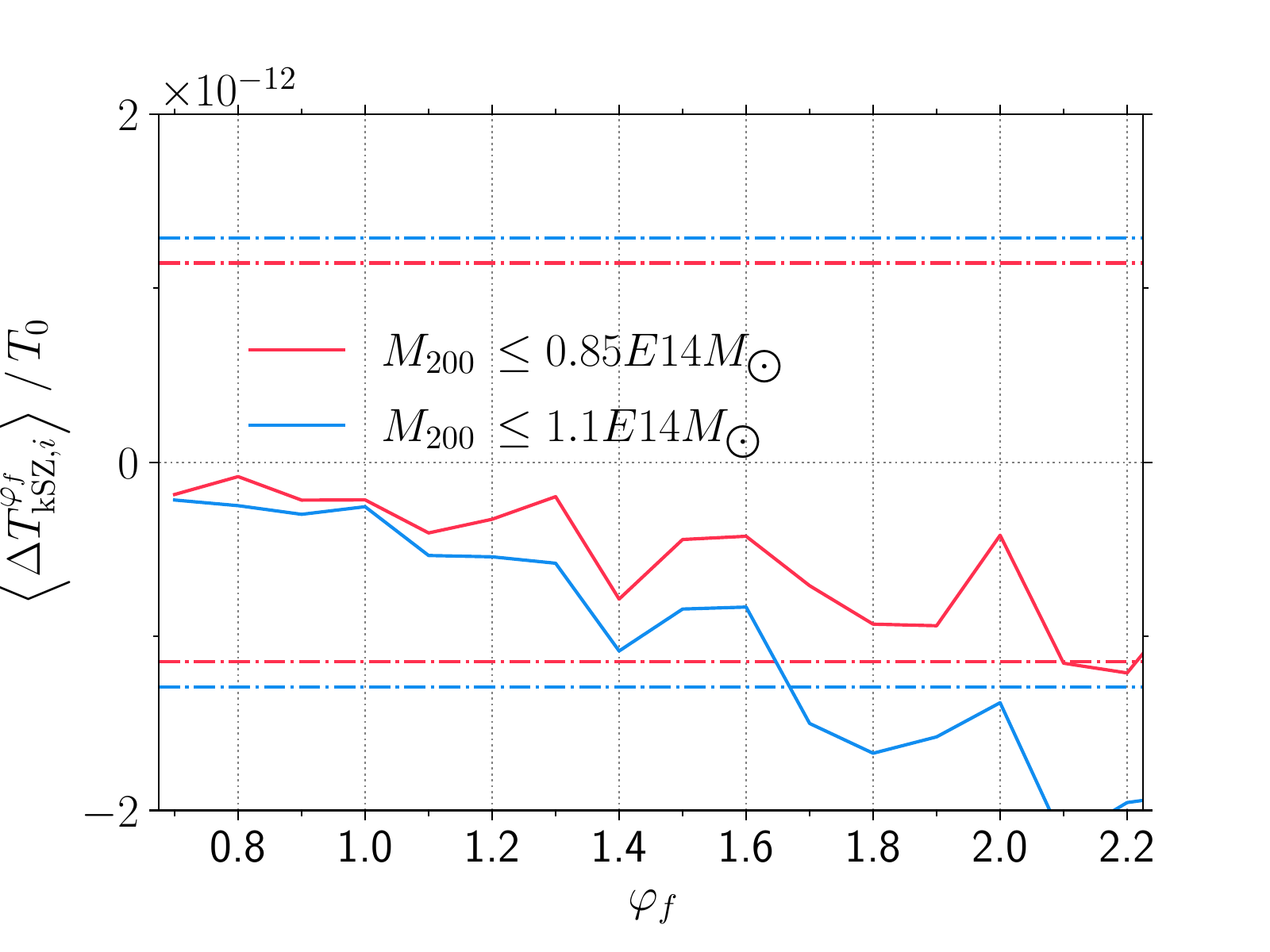}
        }}
    \caption{The average AAP filter output from \smica2018 CMB map at locations of maxBCG clusters below $1.1\times10^{14}\Msun$, binned in $M_{200}$ (left panel) and cumulative (right panel). The dot-dashed lines represent the typical kSZ signal amplitude of clusters in the $M_{200}$ bin of the same color, assuming a universal $v^{\LOS}=300\,\mathrm{km} \mathrm{s}^{-1}$.}\label{fig:maxBCG_M200cut}
\end{figure*}

   Ref.~\cite{Schaan:2016} pointed out that the tSZ contamination becomes important for rare, massive clusters with total mass larger than $10^{14}\Msun$. We have found that, specifically for our sub-sample of maxBCG clusters after redshift and survey mask cuts and the \smica2018 CMB map, it is necessary to remove all clusters with mass greater than $8.5\times10^{13}\Msun$. This is demonstrated in \reffig{maxBCG_M200cut}, where we show the average AAP filter output, as a function of filter scale, at locations of clusters divided in two equi-log $M_{200}$ bins, up to $M_{200}=1.1\times10^{14}\Msun$. We especially check for the cancellation of tSZ signal (and other possible foreground contaminations) by comparing the average AAP filter output to the typical amplitude of the kSZ signal of clusters in each corresponding mass bin. For simplification, we assume a LOS velocity $v^{\LOS}=300\,\mathrm{km} \mathrm{s}^{-1}$ for all clusters. Both panels of \reffig{maxBCG_M200cut} show signs of a significant tSZ contamination when including clusters more massive than $8.5\times10^{13}\Msun$.

   \section{The kSZ likelihood: mixture weights, MAP mean and variance of the kSZ signal amplitude}
   \label{app:kSZ:gaussianmixture}

   Here, we first derive the individual mixture weight $\lambda_s$ for each component of the Gaussian mixture distribution of large-scale bulk-flow amplitude $\alpha_s$ in \refsec{kSZ:datamodel}.
   Let us denote
   \be
   A_i = \D T_{\kSZ, i}/T_0,
   \ee
   and
   \be
   B_i^s = -\tau_i v^{\LOS, s}_{L, i}/c.
   \ee
   Then, the likelihood on the r.h.s. of \refeq{marginal_posterior} can be re-written as
   \be
   \ln\mathcal{P}\left( \{A_i\}|\alpha,\, \{B_i^s\} \right) = -\frac{1}{2} \sum_i \left[\frac{A_i}{\sigma_i} \right]^2 - \frac{1}{2}\left[\sum_i    \left(\frac{B_i^s}{\sigma_i} \right)^2 \right]\,\left[\alpha^2- 2 \alpha \frac{\sum_i \frac{A_i\, B_i^s}{\sigma_i^2}}{\sum_i    \left(\frac{B_i^s}{\sigma_i} \right)^2 }  \right] -\frac{1}{2} \sum_i \ln\left[\sigma_i^2\right].
   \label{eq:kSZ_ln_likelihood}
   \ee
   We further denote
   \be
   \mu_s = \frac{\sum_i \left[ \left(A_i\, B_i^s\right)/\sigma_i^2 \right]}{\sum_i \left( B_i^s/\sigma_i \right)^2},
   \ee
   \be
   \sigma_s^2 = 1 / \left[ \sum_i \left( \frac{B_i^s}{\sigma_i} \right)^2 \right],
   \ee
   \be
   \gamma = \sum_i \left(\frac{A_i}{\sigma_i} \right)^2,
   \ee
   and
   \be
   \delta= \sum_i \ln\left[\sigma_i^2\right],
   \ee
   so that we can shorten \refeq{kSZ_ln_likelihood} to
   \be
   \ln\P = -\frac{1}{2} \gamma + \frac{\mu_s^2}{2\,\sigma_s^2} - \frac{1}{2} \delta - \frac{1}{2\,\sigma_s^2}\,\left(\alpha-\mu_s\right)^2.
   \label{eq:kSZ_ln_likelihood_short}
   \ee
   The first and third terms on the r.h.s. of \refeq{kSZ_ln_likelihood_short} neither vary between \borg-SDSS3 samples nor depend on $\alpha$, thus they can be dropped from \refeq{kSZ_ln_likelihood_short}, such that
   \be
   \ln\P \propto \frac{1}{2\,\sigma_s^2} \left[ \mu_s^2 - \left(\alpha-\mu_s  \right)^2 \right].
   \label{eq:logP}
   \ee
   We can now simply write
   \be
   \P\left(\alpha|\{\D T_{\rm kSZ,\it i}/T_0\}, \{\tau_i v^{\rm LOS,\it s}_{L,\it i}/c\} \right)  \propto  \mathcal{P}(\alpha) \frac{1}{N} \sum_s^N \mathrm{e}^{\omega_s}\, \sqrt{2\pi\,\sigma_s^2 } \frac{\exp\left({-\,\frac{\left(\alpha-\mu_s\right)^2}{\,  2\sigma_s^2}}\right)}{\sqrt{2\pi\,\sigma_s^2 }}
   \ee
   where $\omega_s\equiv\frac{\mu^2_s}{2\sigma^2_s}$.
   The \emph{normalized} version will then be
   \bab
   \P\left(\alpha|\{\D T_{\rm kSZ,\it i}/T_0\}, \{\tau_i v^{\rm LOS,\it s}_{L,\it i}/c\} \right) &=  \mathcal{P}(\alpha) \frac{\frac{1}{N} \sum_s^N \mathrm{e}^{\omega_s }\, \sqrt{2\pi\,\left(\sigma_s\right)^2 } \frac{\exp\left(-\frac{\left(\alpha-\mu_s\right)^2}{\,  2\left(\sigma_s\right)^2}\right)}{\sqrt{2\pi\,\left(\sigma_s\right)^2 }}}{\frac{1}{N} \sum_s^N \mathrm{e}^{\omega_s }\, \sqrt{2\pi\,\left(\sigma_s\right)^2 } } \\
   &=  \mathcal{P}(\alpha)  \sum_s^N \lambda_s \frac{\exp\left(-\,\frac{\left(\alpha-\mu_s\right)^2}{\,2\left(\sigma_s\right)^2}\right)}{\sqrt{2\pi\,\left(\sigma_s\right)^2 }},
   \label{eq:weighted_mixture_gaussian_PDF}
   \eab
   where
   \be
   \lambda_s = \frac{\mathrm{e}^{\omega_s \, + \frac{1}{2}\ln\left[2\pi\,\left(\sigma_s\right)^2 \right]}}{ \sum_s^N \mathrm{e}^{\omega_s \, + \frac{1}{2}\ln\left[2\pi\,\left(\sigma_s\right)^2 \right]}}.
   \ee
   The mean of this distribution can be derived as follows, assuming a uniform prior, i.e. $\P(\alpha)=1$,
   \bab
   \<\alpha\>_s &= \int d\alpha \, \alpha \, \sum_s^N \lambda_s \, \frac{\exp\left(-\,\frac{\left(\alpha-\mu_s\right)^2}{\,2\left(\sigma_s\right)^2}\right)}{\sqrt{2\pi\,\left(\sigma_s\right)^2 }} \\
   &= \sum_s^N \lambda_s \, \int d\alpha \, \alpha \, \frac{\exp\left(-\,\frac{\left(\alpha-\mu_s\right)^2}{\,2\left(\sigma_s\right)^2}\right)}{\sqrt{2\pi\,\left(\sigma_s\right)^2 }} = \sum_s^N \lambda_s \mu_s,
   \eab
   which is precisely \refeq{alpha_ensemble_mean}.

   Similarly, we can explicitly work out the variance of the Gaussian mixture distribution.
   \bab
   \sigma^2_{\alpha} &= \< \left( \alpha -\< \alpha \>_s\right)^2 \> \\
   &=\int d\alpha\,\mathcal{P}\left(\alpha|\{\D T_{\kSZ, i}/T_0\}, \{\tau_i v^{\LOS, s}_{L, i}/c\} \right) \, \left( \alpha -\< \alpha \>_s\right)^2 \\
   &= \sum_s^N \lambda_s  \int d\alpha\, \frac{\exp\left(-\,\frac{\left(\alpha - \mu_s\right)^2}{\,2\left(\sigma_s\right)^2}\right)}{\sqrt{2\pi\,\left(\sigma_s\right)^2 }} \, \left( \alpha -\< \alpha \>_s\right)^2 \\
   &=  \sum_s^N \lambda_s  \left(\sigma_s\right)^2  + \sum_s^N \lambda_s \,\left[ \left(\mu_s - \langle \alpha \rangle_s\right)^2\right].
   \label{eq:MAP_variance}
   \eab
   So the result is the sum of the average noise variances and the variance of the mean estimate (cf. \refeq{alpha_ensemble_variance}). The second term clearly shows that our uncertainty on $\alpha$ includes also the uncertainty from the velocity reconstruction.

   We next compute $\ln\mathcal{P}\left(\{\v{A_i}\} |\alpha , \{\v{B^s_i}\} \right)$ for measurements at all $\theta_f$ scales in a similar fashion to how we arrived at \refeq{weighted_mixture_gaussian_PDF}:
   \bab
   \ln\P\left(\{\v{A_i}\} |\alpha , \{\v{B^s_i}\} \right) &= -\frac{1}{2}\sum_i \,\left[\v{A_i} - \alpha \v{B_i^s}\right]^\intercal \,\left(\v{C}_i\right)^{-1} \,\left[\v{A_i} - \alpha \v{B_i^s}\right] - \frac{1}{2}\sum_i \ln |\v{C}_i| \\
   &\propto -\frac{1}{2}\sum_i \left(\v{B_i^s}\right)^\intercal \,\left(\v{C_i}\right)^{-1} \,\v{B_i^s} \left[\alpha^2 - 2\alpha\frac{\sum_i \v{A_i}^\intercal \,\left(\v{C_i}\right)^{-1} \,\v{B_i^s}}{\sum_i \left(\v{B_i^s}\right)^\intercal \,\left(\v{C_i}\right)^{-1} \,\v{B_i^s}}\right]
   \label{eq:combined_logP}
   \eab
   where we have omitted terms that do not vary between \borg-SDSS3 samples.
   \refeq{combined_logP} is similar to \refeq{logP} with
   \be
   \mu_s = \frac{\sum_i \v{A_i}^\intercal \,\left(\v{C_i}\right)^{-1} \,\v{B_i^s}}{\sum_i \left(\v{B_i^s}\right)^\intercal \,\left(\v{C_i}\right)^{-1} \,\v{B_i^s}}
   \ee
   and
   \be
   \sigma_s^2= 1 / \left[ \sum_i \left(\v{B_i^s}\right)^\intercal \,\left(\v{C_i}\right)^{-1} \,\v{B_i^s} \right].
   \ee


   \section{CMB contribution to covariance matrix of combined kSZ measurement}
   \label{app:kSZ:AP_covariance}

   We provide here a detailed derivation of the CMB covariance matrix term (cf. \refeq{Fourier_covariance_matrix}) that contributes to the covariance matrix of the combined kSZ measurement described in \refsec{kSZ:multi_scale_model}.
   Let us plug \refeq{cmbap_Fourier} into the first term on the r.h.s. of \refeq{covariance_matrix}, to obtain
   \bab
   \C^{\theta_f\theta_f'}_{\CMB,i} &= \left(\frac{\pi\theta_f\theta'_f}{T_0}\right)^2 \left\langle \int \frac{d\vl}{(2\pi)^2} \, \int \frac{d\vl'}{(2\pi)^2}\right.\\
   &\qquad\qquad\qquad\qquad\left.\exp{\left[i\left(\vl-\vl'\right)\cdot\v{\theta_i}\right]} \, \hat{W}\left(\l\theta_f\right) \, \D \Tcmb^{\obs}(\vl) \, \hat{W}\left(\l'\theta'_f\right) \, \D \Tcmb^{\obs}(\vl')\right\rangle \\
   &= \frac{\pi\theta_f^2(\theta'_f)^2}{2T_0^2}\,\int_0^\infty d\l \, \l \, \hat{W}\left(\l\theta_f\right)\, \hat{W}\left(\l\theta_f'\right) \, \Cl{\CMB}.
   \label{eq:Fourier_covariance_matrix_derivation}
   \eab

   \begin{figure}
    \centerline{\resizebox{0.7\textwidth}{!}{\includegraphics*{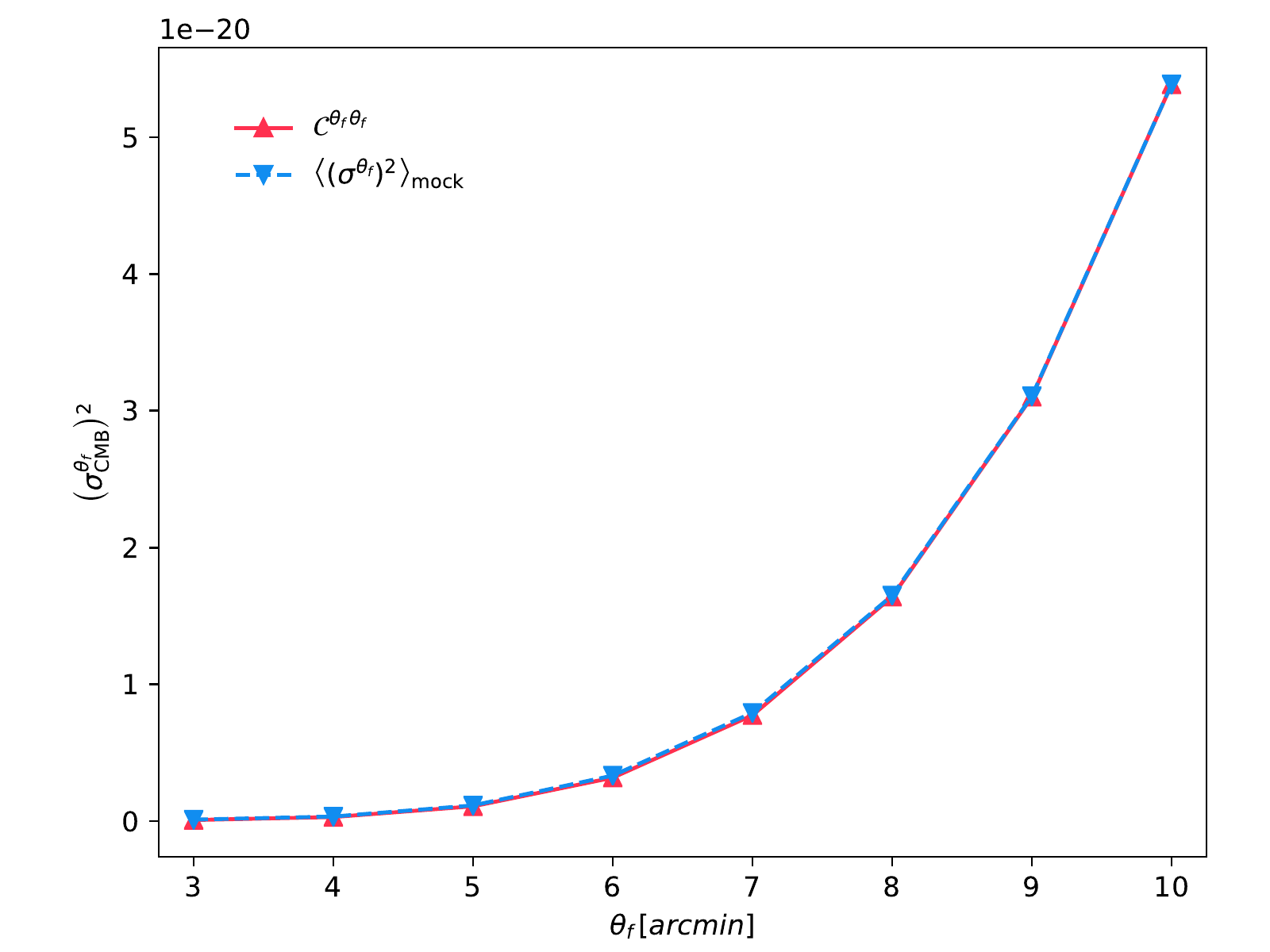}
    }}
    \caption{Comparison between analytical (red) and numerical (blue) estimates of CMB contribution to the diagonal of combined kSZ signal covariance matrix.}\label{fig:theoretical_vs_numerical_sigmai_SMICA2018}
  \end{figure}

   We further validate the analytical computation of the CMB covariance matrix in \refeq{Fourier_covariance_matrix_derivation} by comparing its diagonal elements with the sample variance of $\D T_{\CMB}^{\theta_f}(\v{\theta_i})/T_0)$ computed at 1000 random points in each of the 1000 \smica2018-like CMB mocks in \reffig{theoretical_vs_numerical_sigmai_SMICA2018}.
   The analytical calculatrion using the Planck 2018 best-fit $\L$CDM power spectrum is in extremely good agreement with the numerical estimate using \smica2018-like CMB realizations.


   \section{GADGET-2 simulation of the \borg-SDSS3 volume and mock kSZ signal templates}
   \label{app:kSZ:mock}

\begin{figure}
    \centerline{\resizebox{0.7\textwidth}{!}{\includegraphics*{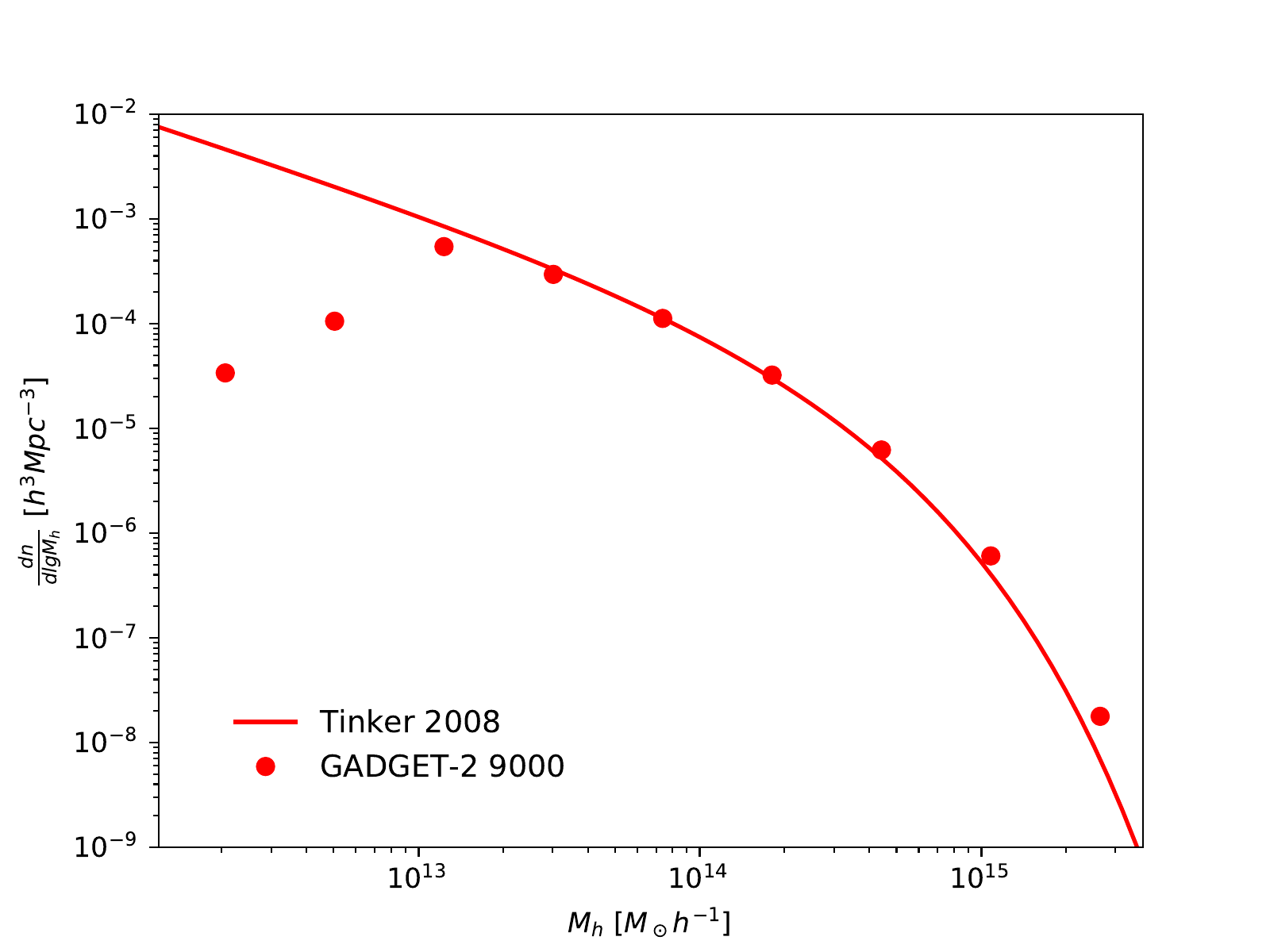}
    }}
    \caption{The number density of DM halos identified by Rockstar in our high-resolution GADGET-2 simulation of \borg-SDSS3 sample 9000, as compared to the Tinker 2008 halo mass function \cite{Tinker:2008}, at redshift $z=0.23$.}\label{fig:resim}
  \end{figure}

   In order to generate the mock template of kSZ signal within the \borg-SDSS3 volume, we use DM particles and halos from a GADGET-2 \cite{Springel:2005} simulation with DM-only at a very high resolution of $N_{\mathrm{part}}=2048^3$. The initial conditions for the simulation is taken from the \borg-SDSS3 sample $s=9000$. The halos are identified as main halos by the Rockstar halo finder algorithm\footnote{\url{https://bitbucket.org/gfcstanford/rockstar}} \cite{Behroozi:2013, Knebe:2013} with a minimum number of 20 particles per halo.
   The cosmology and box size of this simulation agree with those of our \borg-SDSS3 reconstruction; the initial conditions are specifically taken from sample 9000. The high resolution allows us to achieve a correct halo mass function (HMF) down to $M_{h}=2\times10^{13}\Msunh$ at redshift $z=0.23$, as shown in \reffig{resim}.
  
For the validity test of our estimator, we model the gas profile of each halo $i$ with a Gaussian profile \citep[see, e.g.][]{Schaan:2016,Sugiyama:2018}:
   \be
   n_{e,i}(\theta) = \frac{N_{e,i}}{\sqrt{2\pi\theta^2_i}}\exp\left(-\frac{\theta^2}{2\theta^2_i}\right)
   \ee
   where $\theta^2_i=\theta^2_{200,i}+\theta^2_{\beam}$ and $N_{e,i}=(f_b M_{200,i})/(\mu_e m_p)$.
For the measurement of the signal profile in \refsec{kSZ:multi_scale_model}, we directly generate the kSZ template using all individual DM particles within the same volume analyzed in this work.
The LOS velocity of each particle (or halo) is interpolated from the \tcola{} simulation of \borg-SDSS3 sample 9000.


\bibliographystyle{JHEP}
\bibliography{references}

\end{document}